\documentclass[openacc]{rsproca_new}

\definecolor{DarkGreen}{rgb}{0.0,0.45,0.0}     
\definecolor{DarkMagenta}{rgb}{0.45,0.0,0.45}  

\usepackage{graphicx}

\begin{document}

\title{Magnetic reconnection in partially ionized plasmas}

\author{
Lei Ni$^{1,5}$, Hantao Ji$^{2,3}$, Nicholas A. Murphy$^{4}$ and Jonathan Jara-Almonte$^{3}$}

\address{$^{1}$Yunnan Observatories, Chinese Academy of Sciences, P. O. Box 110, Kunming, Yunnan 650216, P. R. China\\
$^{2}$Department of Astrophysical Sciences, Princeton University, Princeton, New Jersey 08544, USA\\
$^{3}$Princeton Plasma Physics Laboratory, Princeton, New Jersey 08543, USA\\
$^{4}$Center for Astrophysics $\vert$ Harvard \& Smithsonian, 60 Garden Street, Cambridge, Massachusetts 02138, USA\\
$^{5}$Center for Astronomical Mega-Science, Chinese Academy of Sciences, 20A Datun Road, Chaoyang District, Beijing 100012, P. R. China}

\subject{plasma physics, solar physics, astrophysics}

\keywords{magnetic reconnection, xxxx, xxxx}

\corres{Lei Ni\\
\email{leini@ynao.ac.cn}}

\begin{abstract}
Magnetic reconnection has been intensively studied in fully ionized plasmas. However, plasmas are often partially ionized in astrophysical environments. The interactions between the neutral particles and ionized plasmas might strongly affect the reconnection mechanisms. We review magnetic reconnection in partially ionized plasmas in different environments from theoretical, numerical, observational and experimental points of view. We focus on mechanisms which make magnetic reconnection fast enough to compare with observations, especially on the reconnection events in the low solar atmosphere. The heating mechanisms and the related observational evidence of the reconnection process in the partially ionized low solar atmosphere are also discussed. We describe magnetic reconnection in weakly ionized astrophysical environments, including the interstellar medium and protostellar disks. We present recent achievements about fast reconnection in laboratory experiments for partially ionized plasmas.
\end{abstract}


\begin{fmtext}

\end{fmtext}

\maketitle

\section{Introduction\label{introduction}}

Magnetic reconnection is a fundamental and important process in many different plasma environments including laboratory experimental devices, planetary magnetospheres, stellar atmospheres, the interstellar medium (ISM), and other astrophysical settings. Magnetic energy is converted into plasma kinetic energy, thermal energy, and acceleration of particles during magnetic reconnection. Distinctly different from simple magnetic diffusion, events involving magnetic reconnection are impulsive in nature by generating Alfv\'enic plasma flows which subsequently dissipate into heat. The global magnetic topology is not necessarily changed during the magnetic diffusion process, while it must be changed by the magnetic reconnection process. More importantly, magnetic diffusion contributed by the pure ohmic diffusion does not typically result in fast magnetic energy release in weakly collisional or collisionless plasmas. The observed rapid energy release events occurring in the Earth's magnetosphere, the solar atmosphere and the other astrophysical environments need to be explained by fast magnetic reconnection. Several review papers~\cite{zweibel:2009,yamada:2010,zweibel:2016} have been published in recent years on the topic of magnetic reconnection in general.  

In fully ionized plasmas, magnetic reconnection has been intensively studied. The magnetohydrodynamic (MHD) model is usually applied to study large-scale reconnection events in solar corona\cite{priest:2000} such as solar flares and coronal jets. In the ideal MHD approximation, the magnetic field is frozen-in to the plasma. Magnetic reconnection breaks the frozen-in condition in the diffusion region on time scales much shorter than the plasma's classical diffusion time based on electron-ion collisions\cite{zweibel:2009}. Plasma is weakly collisional in the solar corona, and collisionless in earth's magnetosphere. The diffusion region around one reconnection X-point in such weakly collisional and collisionless plasmas has a characteristic length scale of about the ion inertial length $r_i$\cite{ren:2005,yamada:2006} or the ion Larmor radius $r_s$\cite{egedal:2007,katz:2010}. Kinetic effects play important roles in causing fast reconnection and dissipating magnetic energy in these plasmas. However, the length scale of a solar coronal flare is about $10 \sim 100$~Mm. The ion inertial length is $r_i=c/\omega_{pi} \sim 200 $~cm. How does the global MHD scale connect to such a small scale diffusion region to cause fast reconnection and the related eruptive event to happen? It is suggested that reconnection before solar coronal eruptive events is collisional, and the dynamics cause the layer to thin to kinetic scales\cite{cassak:2005,cassak:2012}, where kinetic reconnection onsets abruptly.  

Recent numerical and analytical results suggest that magnetic reconnection in solar applications processes through a hierarchy of current sheets and magnetic islands as originally proposed by Shibata and Tanuma \cite{shibata:2001}. Such an unstable reconnection process is also called plasmoid instability\cite{biskamp:2000, bhattacharjee:2009} or tearing instability\cite{furth:1963}. A magnetic island in two dimensional (2D) simulations and analysis is also called a plasmoid and possibly corresponding to the bright blob from observations. Such a magnetic island in 2D is actually a magnetic flux rope in 3D\cite{mei:2017}. The bright blobs observed in flare current sheets \cite{asai:2004, lin:2005,takasao:2012}, jets \cite{zhang:2016, li:2019, shen:2019} and other solar reconnection events \cite{li:2016} may correspond to plasmoids from this instability. The Interface Region Imaging Spectrograph (IRIS) has recently observed sub-arcsecond bright blobs \cite{zhang:2019}, which suggests that smaller and smaller plasmoids will be observed as the resolution of solar telescopes continues to improve.  On the MHD scale, plasmoid instability already leads to a fast reconnection rate of about $\sim 0.01$ \cite{bhattacharjee:2009, huang:2010, ni:2015}, which is in the range measured from observations (about $0.001$ to $0.1$)\cite{lin:2005, yokoyama:2001, isobe:2002, narukage:2006} but about 10 times slower than the reconnection rate measured in kinetic regimes ($\sim0.1$)\cite{shay:1998, hesse:1999, birn:2001, huba:2004, guo:2014}. Particle-in-cell (PIC) simulation results also show that the elongated electron layers with length scales of about the ion inertial length are also generally unstable to plasmoid formation\cite{daughton:2006, karimabadi:2007, klimas:2008}. Fully kinetic PIC simulations with a Monte-Carlo treatment of the Fokker-Plank collision operator have recovered the Sweet-Parker scaling and the transition to the faster kinetic regime\cite{daughton:2009a, daughton:2009b}. These results suggest that the plasmoid instability in large systems may naturally push the evolution towards small scales where kinetic effects become essential in the reconnection process\cite{daughton:2009b}. However, high Lundquist number MHD simulations and large-scale kinetic simulations remain challenging to perform given limits on computing power even in two dimensions. 

The Reynolds number in many astrophysical environment is large enough to result turbulence to appear. The importance of turbulence on magnetic reconnection has been recognized long time ago\cite{matthaeus:1985, matthaeus:1986}. The extension of the Sweet-Parker model of laminar reconnection taking into account the presence of turbulence in MHD scale has been suggested in a implicitly three-dimensional model\cite{lazarian:1999}, which is called LV99 model by the authors. The outflow thickness is then determined by the large-scale magnetic field wandering. The reconnection rate in LV99 model does not prescribe a given value and it is determined by the level of MHD turbulence\cite{lazarian:1999}. The role of turbulence in reconnection was then numerically studied \cite{kowal:2009, kowal:2012}, and the interplay between the turbulence and reconnection and the breakdown of flux-freezing condition at all scales by turbulence was also investigated \cite{eyink:2013}. The recent review paper has summarized the developments of this turbulent reconnection model and compared it with other fast reconnection models\cite{lazarian:2020}.  Although fractal structures and non-thermal spectral broadening have been observed by solar telescopes \cite{georgoulis:2005, jeffrey:2018}, it is unclear whether turbulent reconnection does occur as the proposed LV99 model.

Magnetic reconnection rate is an important quantity that represents the speed of magnetic energy dissipation during the reconnection process. The normalized magnetic reconnection rate mentioned in this review can be measured in several different ways. It is usually defined as $\frac{1}{B_{up} v_A^{\ast}}\frac{d \psi}{dt}$ \cite{bhattacharjee:2009, huang:2010, ni:2015}, where $\psi$ is the magnetic flux at the reconnection region, $B_{up}$ and $v_A^{\ast}$ are the strength of the {\it reconnecting component} of magnetic field and the associated Alfv\'en speed at the reconnection upstream region respectively. It can also be measured as $v_{in}/v_A^{\ast}$ \cite{shen:2011}, where $v_{in}$ is the average inflow velocity, the Alfv\'en speed $v_A^{\ast}$ at the reconnection upstream region is replaced by the reconnection outflow velocity $v_{out}$ in observations of current sheets in solar atmosphere \cite{yokoyama:2001, isobe:2002, lin:2005 }. The reconnection electric field can be measured to calculate the reconnection rate in the PIC simulations \cite{shay:1998, hesse:1999, guo:2014}, experiments \cite{yamada:2000, yamada:2006, lawrence:2013} and observations from earth's magnetosphere\cite{mozer:2002}. In MHD simulations, the dimensionless reconnection rate is normally estimated as $\eta^{\ast} j_X/(B_{up} v_A^{\ast})$ \cite{leake:2012, leake:2013, ni:2018a, ni:2018b, ni:2018c}, where $\eta^{\ast}$ and $j_X$ are the magnetic diffusion coefficient and the current density at the reconnection X-point respectively. These different methods are fundamentally the same under some conditions. In this review, the normalized reconnection rate in Figure \ref{fig8} is calculated as $\gamma=\frac{1}{B_{up} v_A^{\ast}}\frac{d \psi}{dt}$, and it is defined as $M_{sim}=\eta^{\ast} j_X/(B_{up} v_A^{\ast})$ in Figure \ref{fig12} and Figure \ref{fig15}. The Lundquist number $S$ is another important dimensionless parameter, which is defined as $S=v_A^{\ast}L/\eta^{\ast}$, $L$ is the half length of the current sheet. Plasmoid instability occurs only when the Lundquist number exceeds the critical value \cite{biskamp:2000, bhattacharjee:2009, ni:2010}. 

There are substantially fewer works about magnetic reconnection in partially ionized plasmas. In fact, many astrophysical environments are filled with partially ionized plasmas \cite{ballester:2018}, such as the low solar atmosphere, cometary tails, protoplanetary nebulae, disks around young stellar objects, and the ISM\@. Plasmas in many of these environments are weakly ionized. The ionization fraction in the warm neutral medium is about $0.01$, $10^{-7}$ in molecular clouds, and can be as low as $10^{-10}$ in protostellar disks \cite{zweibel:2009}.  Magnetic reconnection plays an important role in the formation and evolution of the large scale magnetic fields in the partially ionized ISM and is also crucial to understand the observed small scale solar activities in the low solar atmosphere. The interactions between ions and neutrals may strongly affect the reconnection process and make the reconnection mechanisms to be very different from those in fully ionized plasmas. The present review article focuses mainly on three topics: (1) How do the neutrals affect the reconnection process from different effects? (2) What is the dominant mechanism to lead the fast reconnection and under what kind of conditions can the fast reconnection set in in different astrophysical environment with partially ionized plasmas? (3) What do the present observations from telescopes and laboratory experiments tell us about magnetic reconnection in partially ionized plasmas? Can the present theoretical models and numerical results explain these observed events? 

This paper is organized as follows. In \S\ref{theories}, we introduce theories on magnetic reconnection in partially ionized plasmas; discuss different aspects of how neutrals affect the reconnection process; and describe situations when the neutrals and ions will be strongly coupled, weakly coupled, or intermediately coupled.We discuss magnetic reconnection in the low solar atmosphere from observational, theoretical, and numerical perspectives in \S\ref{solar}.In \S\ref{astrophysics} and \S\ref{experiments}, we summarize investigations of magnetic reconnection in other astrophysical environments and laboratory experiments, respectively. 
In \S\ref{summary}, we discuss what we have learned and currently understand about magnetic reconnection in partially ionized plasmas from the recent developments summarized here, as well as remaining open questions and perspectives on the future.

\section{The basic theories about how the neutrals affect magnetic reconnection\label{theories}}
Neutrals can affect magnetic reconnection process in several different ways. The effect of neutrals on magnetic diffusivity is the most obvious and intuitive one. When including the collisions between electrons and neutrals, the magnetic diffusion coefficient is 
\begin{align}\label{2.1}
\begin{split}
\eta=\frac{m_e(\nu_{ei}+\nu_{en})}{e^2n_e\mu_{0}},
\end{split}
\end{align} 
and the corresponding electron-ion ($\nu_{ei}$) and electron-neutral ($\nu_{en}$) collisional frequencies are :
\begin{align}\label{2.2}
\begin{split}
\nu_{ei}=\frac{4}{3}n_i\sigma_{ei}\sqrt{\frac{2k_BT_{ei}}{\pi m_{ei}}}, \nu_{en}=n_n \sigma_{en} \sqrt{\frac{8k_BT_{en}}{\pi m_{en}}}
\end{split}
\end{align} 
where the particle mass $m_{\alpha \beta}=m_{\alpha}m_{\beta}/(m_{\alpha}+m_{\beta})$, the plasma temperature $T_{\alpha \beta}=(T_{\alpha}+T_{\beta})/2$, $\alpha$ and $\beta$ represent the different species of particles. $n_i$ and $n_n$ are the number density of ions and neutrals respectively. $\mu_0$ is the magnetic permeability coefficient of free space. $k_B$ is the Boltzmann constant. $\sigma_{en}$ is the electron-neutral collisional cross section \cite{draine:1983} which can usually be assumed as a constant \cite{khomenko:2012, leake:2013}. The electron-ion collisional cross-section depends on the plasma temperature as $\sigma_{ei}=\Lambda \pi r_{d,e}^2$, where $\Lambda$ is the Coulomb logarithm, $r_{d,e}=e^2/(4 \pi \epsilon_0 k_B T_e)$, $e$ is the elementary charge, $\epsilon_0$ is the permittivity of free space.  In some astrophysical environments such as protostellar disks, $\nu_{en}/\nu_{ei}$ is about $10^4-10^7$\cite{malyshkin:2011}. Therefore, electron-neutral collisions dominate the magnetic diffusion coefficient and it is very important in the magnetic reconnection process in such environments with a Lundquist number from about $10^3$ to $10^8$\cite{malyshkin:2011}. However, one should keep in mind that magnetic reconnection becomes meaningless if the collisional frequencies are too large and the corresponding Lundquist number of the system is too low. For example, in protoplanetary disks, the combination of the high magnetic diffusion coefficient and the relatively small size of the system makes the Lundquist number $S$ fall below unity\cite{zweibel:2009, hayashi:1981}, and pure Ohmic dissipation instead of magnetic reconnection will dissipate magnetic fields. Therefore, it is necessary to do the analysis and to find out if the system to be studied is in the intermediate regime, in which electron-neutral collisions are important and the corresponding Lundquist number is also large enough for triggering magnetic reconnection\cite{wardle:1999}. 

Reconnection occurs on a time scale which is intermediate between the resistive time scale and the much faster Alfv\'en time scale. In partially ionized plasmas, the neutrals might strongly, intermediately or weakly couple with the ionized plasmas depending on the frictions between them\cite{zweibel:1989}. The ionized plasmas and neutrals are considered to move together when the perturbations are on timescales much longer than the ion-neutral collision time $\tau_{in}=(\rho_i/\rho_n)\tau_{ni}$, the Alfv\'en speed depends on the combined mass density of both species for such a strongly coupling case, and it is defined as  $v_A=B/\sqrt{\mu_0(\rho_i+\rho_n)}$, where $B$ is the strength of the magnetic field, $\rho_i$ and $\rho_n$ are the ion density and neutral density respectively. When the perturbations are on timescales much shorter than the ion-neutral collisions time $\tau_{in}$, the ions are then decoupled from neutrals and the Alfv\'en speed only depends on the ionized components as $v_{Ai}=B/\sqrt{\mu_0\rho_i}$.  In weakly ionized plasmas, $v_{Ai}$ could be several orders of magnitude higher than $v_{A}$, the Alfv\'en speed is reduced by a factor of $\sqrt{\rho_n/\rho_i}$ in the strongly coupling case. Both the steady state magnetic reconnection and the time-dependent tearing modes in partially ionized plasmas have been analytically studied\cite{zweibel:1989, zweibel:2011}, the strong, intermediate and weak coupling cases are all considered. In the MHD scale, the weakly coupled reconnection proceeds more rapidly than strongly coupled reconnection by a fractional power of the density ratio $\rho/\rho_i$\cite{zweibel:1989}, where $\rho=\rho_n+\rho_i$. In the context of tearing instabilities in a sheared magnetic field, the reconnection rate is found to be increased by a factor of $(\rho/\rho_i)^{1/5}$ in the weakly coupling case with weakly ionized plasmas\cite{zweibel:1989}.  However, the reconnection rate still follows the $S^{-3/5}$ dependence found in the theory of resistive tearing modes in fully ionized plasmas. In the intermediately coupling case, the dissipation by ion-neutral friction is especially strong and contributes a substantial fraction of the energy dissipated in reconnection\cite{zweibel:2011}.         

When the friction coupling between the ionized plasmas and neutrals is strong, the partially ionized plasmas can then be treated as a single fluid.  However, the extra terms which reflect the ion-neutral drift can be included in both the energy and the magnetic induction equation. These extra terms depend on ambipolar diffusion\cite{brandenburg:1994, brandenburg:1995, zweibel:1997}, which refers to the decoupling of neutrals and charged components. Ambipolar diffusion causes magnetic fields to diffuse through neutral gas due to collisions between neutrals and ionized particles.  Since the neutrals lag behind the magnetic fields, the neutral pressure support is lost\cite{heitsch:2003b}.  In the case with weakly ionized plasmas and zero guide field, the ion pressure which is much smaller than the neutral pressure is overwhelmed by the magnetic pressure\cite{heitsch:2003b}, the ionized plasmas then move toward the neutral sheet, transporting the field lines with it. The magnetic field gradients steepen around the null plane, leading to high current densities and ohmic dissipation in a thin current sheet\cite{brandenburg:1994}. Including the magnetic diffusion, ambipolar diffusion and Hall effect caused by the decoupling of ions and electrons, the magnetic induction equation can then be written as
\begin{align}\label{2.3}
\begin{split}
 \partial_t \mathbf{B}=\nabla \times (\mathbf{v} \times \mathbf{B}-\eta\nabla \times \mathbf{B}+\mathbf{E}_\mathrm{AD}-\eta_H \frac{\nabla \times \mathbf{B} \times \mathbf{B}}{\lvert \mathbf{B} \rvert}),
 \end{split}
\end{align}  
where $\eta_H=\lvert \mathbf{B} \rvert/(en_e\mu_0)$ is the Hall effect coefficient, $\mathbf{E}_\mathrm{AD}$ is the ambipolar diffusion field and it is given by\cite{braginskii:1965}
\begin{align}\label{2.4}
\begin{split}
 \mathbf{E}_\mathrm{AD}=\frac{\eta_{AD}}{{\lvert \mathbf{B} \rvert}^2}[(\nabla \times \mathbf{B})\times \mathbf{B} ] \times \mathbf{B}.
 \end{split}
\end{align}  
The ambipolar diffusion coefficient $\eta_{AD}$ is defined as
\begin{align}\label{2.5}
\begin{split}
 \eta_{AD} = \frac{{\lvert \mathbf{B} \rvert}^2}{\mu_0}\left(\frac{\rho_n}{\rho} \right)^2(\rho_i \nu_{in}+\rho_e\nu_{en})^{-1},
\end{split}
\end{align}
where $\rho_n$, $\rho$, $\rho_i$ and $\rho_e$ represent the neutral, total, ion and electron density, respectively. When $\mathbf{v}=0$ and the field lines are straight with $\mathbf{B}=B(x,t) \hat{e}_z$, the magnetic induction equation in $z$-direction then becomes
\begin{align}\label{2.6}
\begin{split}
  \frac{\partial B}{\partial t}=\frac{\partial}{\partial x}\left(\eta \frac{\partial B}{\partial x}\right)+\frac{\partial}{\partial x} \left( \eta_{AD} \frac{\partial B}{\partial x} \right),
 \end{split}
\end{align}
Equation (\ref{2.6}) shows that in slab geometry, ambipolar drift acts like nonlinear diffusion. In the absence of resistivity, a steady state solution for the magnetic field is $B \propto x^{1/3}$, which has singular current and infinite drift at $x=0$. The numerical results show that the initially linear magnetic profile $B \propto x$ steeps to $B \propto x^{1/3}$ in the reconnection process. The resistivity and the buildup of a plasma pressure gradient might remove this singularity, but these effects are very small in the interstellar medium. One should notice that the additional magnetic field perpendicular to the reconnection plane inhibits current sheet formation\cite{zweibel:1997}, even a small non-zero guide field might drastically reduce the reconnection rate of reversing components\cite{heitsch:2003b}.

The reconnection rate can be strongly enhanced when the recombination times of ions are short \cite{vishniac:1999, heitsch:2003a}. As discussed above, ambipolar diffusion referring to the decoupling of neutrals and charged components leads to a thin current sheet and much faster reconnection rate. The ionized plasmas flow into the resistive layer, leaving the neutrals behind. The recombination effect then reduces the ion pressure in the resistive layer and the current sheet is thinned further. In the classical Sweet-Parker reconnection model, the outflow time is defined as $t_{flow}=L/v_A$ and the inflow velocity can be written as $v_{in} = (\eta/t_{flow})^{1/2}$ \cite{zweibel:2009}. When the resulting excess ions recombine on a timescale $t_{rec}$ which is much shorter than the plasma outflow time, $t_{rec} \ll t_{flow}$,  the reconnection rate can be approximated using $v_{in}\sim (\eta/t_{rec})^{1/2}$ \cite{zweibel:2009, vishniac:1999, heitsch:2003a}. Though the reconnection rate depends on the magnetic diffusion just as in the slow Sweet-Parker model $\sim \eta^{1/2}$, the short recombination time can still make magnetic reconnection process quite fast in some astrophysical environments\cite{heitsch:2003a}. The strong recombination represents a sink, thus the current sheet  can keep in a steady state in one dimension and the plasma outflow is no longer needed in the previous analytical work\cite{heitsch:2003a}. However, the recent 2D numerical simulations show that the plasma outflow and recombination both play important roles in determining the reconnection rate\cite{leake:2012, leake:2013}. The ionization degree and the plasma $\beta$ also affect the reconnection rate\cite{vishniac:1999, heitsch:2003a},  the smaller plasma $\beta$ of ions is expected to make reconnection faster\cite{heitsch:2003a}. The two-dimensional problem gives an analytical result of the reconnection rate \cite{vishniac:1999} in the interstellar medium as $\sim (\eta/t_{rec})^{1/2}(1+2\chi \beta_w)^{1/2}(\chi \beta_w)^{-1}$, where $\chi$ is the ionization fraction and $\beta_w$ is the plasma $\beta$ of the whole plasmas including both ions and neutrals. Then the one dimensional analytical result that accounts for plasma $\beta$ of ions \cite{heitsch:2003a} is given as $\sim (\eta/t_{rec})^{1/2}(\beta_i)^{-3/(2\Gamma)}$, where $\beta_i$ is the plasma $\beta$ of ions and $\Gamma$ is the polytropic exponent by assuming the plasmas in polytropic state. One should keep in mind that this type of reconnection relating to the tendency to form thin current layers in weakly ionized plasmas is suppressed even by a weak guide field\cite{heitsch:2003b}.         

The partially ionized plasmas are usually in non-equilibrium ionization state in a certain time period. When reconnection happens, ohmic heating increases the plasma temperature and ion pressure. The recombination time is then longer, which leads to the lower reconnection rate and a thicker current layer. Thus, the reconnection rate might change with time, depending on the competition between ionization and recombination. The previous analytical work has discussed whether the electron drift associated with current flow is sufficient to ionize the plasmas\cite{heitsch:2003a}, the results show that the equilibrium ionization fraction associated with ionization by electron drift is generally small\cite{heitsch:2003a}. The authors conclude that the reconnection rate depends rather weakly on temperature\cite{heitsch:2003a}, and the effect of the electron drift associated with current flow can be ignored. However, the recent 2D multi-fluid simulations show that the ionization is larger than the recombination in a reconnection process with low $\beta$ in the low solar chromosphere\cite{ni:2018a, ni:2018b, ni:2018c}, the plasmas inside the current sheet can even become fully ionized from initial weakly ionized plasmas\cite{ni:2018a, ni:2018b, ni:2018c}, and the small recombination effect can not result in fast reconnection\cite{ni:2018a}. These previous works only focus on ohmic heating. The dissipation of magnetic energy on kinetic scales in partially ionized plasmas was not considered.

\begin{figure}[h]
 \centering\includegraphics[width=2.2in]{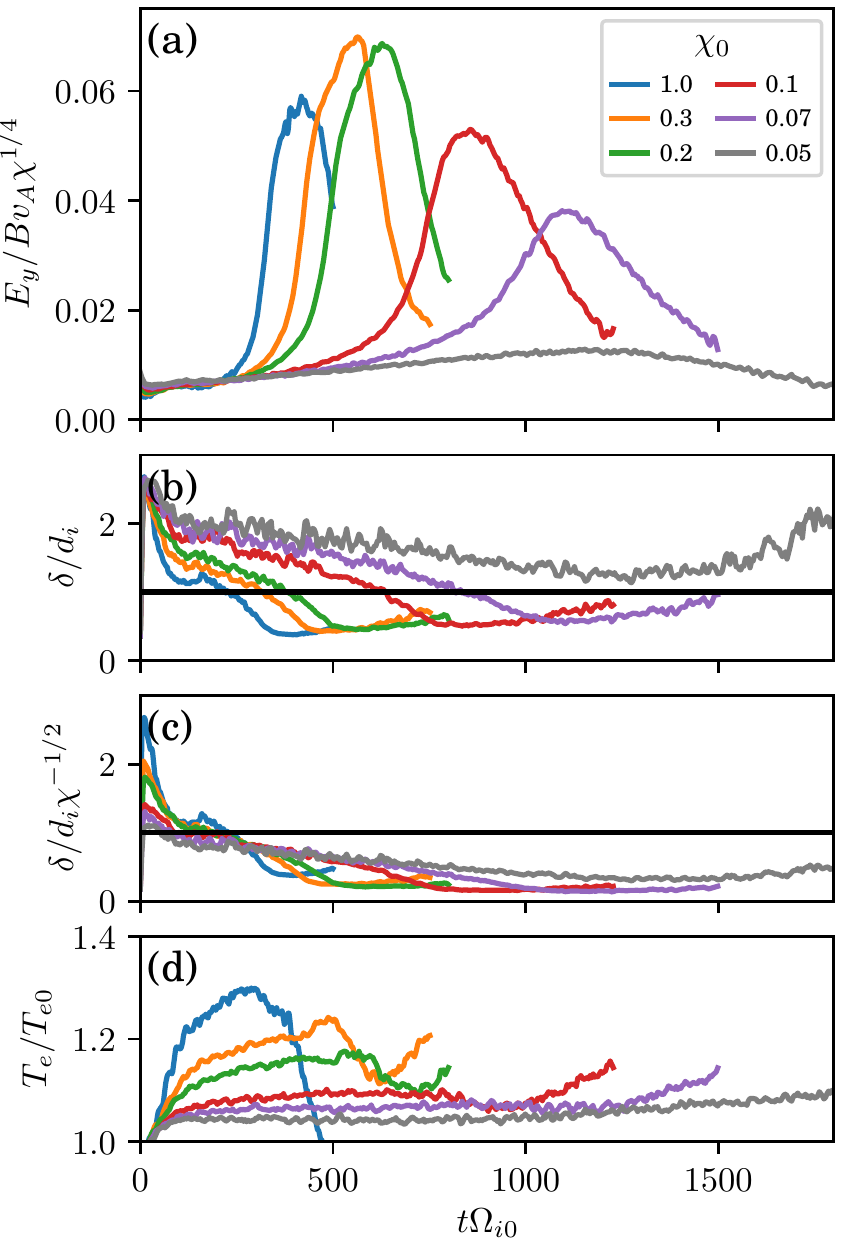}
 \caption{Transition between collisional and collisionless
reconnection for different ionization fraction $\chi$. Panels show (a) the scaled reconnection rate, (b) and (c) the minimum current sheet width normalized to the local ion inertial length $d_i$ and the hybrid inertial length $d_i\chi^{1/2}$, and (d) the electron temperature at the X point. From Ref.\cite{jara:2019}.}
 \label{transition}
\end{figure}

Experiment, theory and simulation for fully ionized plasmas have shown that Hall effect starts to lead fast reconnection when the current sheet thickness is less than the ion inertial length $r_i$\cite{ren:2005,yamada:2006}. In the partially ionized plasmas, the onset of Hall reconnection for the weakly coupling case only depends on the parameters of  ionized plasmas and it is the same as in fully ionized plasmas. In the strongly coupled case with weakly ionized plasmas, the analytical results show that the collisions between neutrals and ions make the effective ion inertial length increase to $r_i \sqrt{\rho_n/\rho_i}$\cite{malyshkin:2011, zweibel:2011}. Thus, the entrainment of neutrals slows MHD reconnection but permits the onset of fast collisionless reconnection at a larger Lundquist number\cite{malyshkin:2011}, and the transition of fast reconnection is expected to happen when the current width $\delta$ reaches in the scale $r_i < \delta < r_i\sqrt{\rho_n/\rho_i}$\cite{malyshkin:2011, zweibel:2011}. Recently, magnetic reconnection in partially ionized plasmas have been studied in experiment devices\cite{lawrence:2013, takahata:2019} (see Sec.5). Laboratory studies of Hall reconnection in partially ionized plasmas show that the width of the ion diffusion layer is not expanded as expected\cite{lawrence:2013}, and the Hall reconnection rate is reduced\cite{lawrence:2013}. The authors pointed out that the size of the experimental device (MRX) is smaller that the predicted effective ion inertial length $r_i \sqrt{\rho_n/\rho_i}$\cite{lawrence:2013}, which is the possible reason for inconsistencies between the analytical result and the experimental result. Nevertheless, the recent first fully kinetic PIC simulations of partially ionized reconnection show that the transition to fast reconnection occurs when the current sheet width thins below the ion inertial length $r_i$\cite{jara:2019}, which is also in contrast to previous analytical predictions. 
Figure~\ref{transition} shows numerical results on the transition to fast reconnection when the current thickness thins below the ion skin depth, rather than the hybrid ion skin depth, for various ionization fractions~\cite{jara:2019}. 
The most illustrative example is the case of $\chi=5\%$ when the current sheet thickness never thins down below the ion skin depth (thus the reconnection stays slow) even if it stays below the hybrid value for most of the time.

The ion-neutral collision and neutral viscosity can damp the MHD turbulence. The previous analytical results\cite{lazarian:2004} found that the turbulent reconnection speeds based on the LV99 model under typical conditions in interstellar medium (ISM) are reduced by the presence of neutrals, but no more than an order of magnitude. Since the neutrals can affect the reconnection rate from several different ways as described above, the plasma and magnetic field parameters should be considered to decide which effect is more important when we study the reconnection event in a specific environment. Furthermore, the dominate fast reconnection mechanism can also possibly change with time during a magnetic reconnection process.

\section{Magnetic reconnection in the low solar atmosphere \label{solar}}
\subsection{The reconnection events from observations}
\begin{figure}[!ht]
 \centering\includegraphics[width=2.2in]{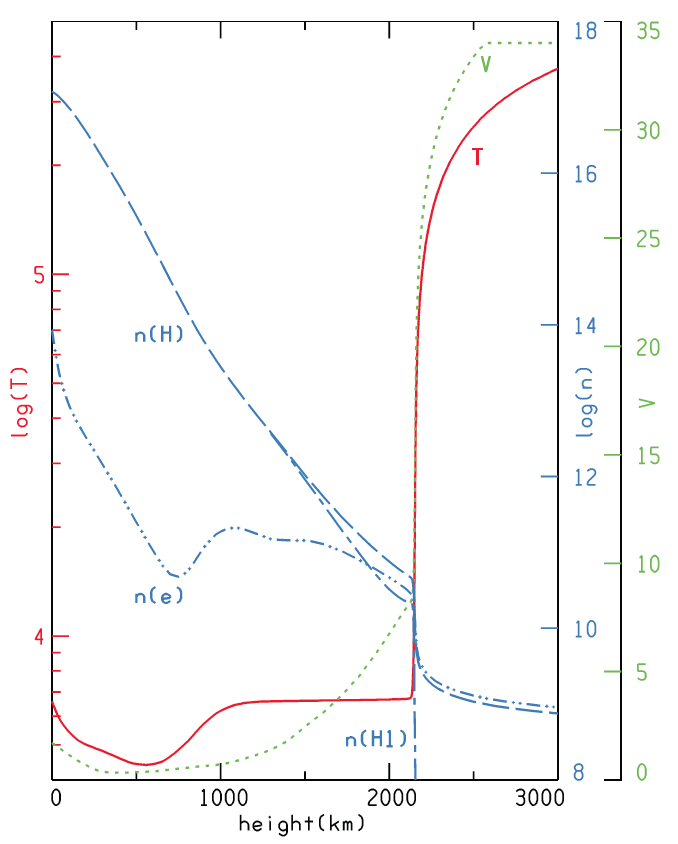}
 \caption{The variations with height in C7 Model, the different lines with different colors and styles represent the temperature $T$ (in K), turbulent velocity $V$ (in km\,s$^{-1}$), total hydrogen density $n(H)$ (in cm$^{-3}$), neutral hydrogen density $n(\mbox{HI})$ (in cm$^{-3}$), and electron density $n(e)$ (in cm$^{-1}$). Image reproduced with permission from Avrett $\&$ Loeser (2008), copyright by AAS.}
 \label{fig2}
\end{figure}

As we know, the temperature of the solar corona is about $1$~MK or even higher and the hydrogen gas which occupies three quarters of the total gas is fully ionized. The plasma characteristics in the low solar atmosphere are totally different from those in the solar corona. According to the solar atmosphere model\cite{vernazza:1981, avrett:2008}, the plasma temperature is about several thousand Kelvin below the upper chromosphere, and the temperature minimum region (TMR) which is located between the upper photosphere and the lower solar chromosphere is only about $4200$~K\@. The plasmas are highly stratified within a thin layer about $2000$~km and the total hydrogen density varies from  $10^{23}$\ m$^{-3}$ at the bottom photosphere to about $10^{16}$\,m$^{-3}$ around the up chromosphere in the quiet sun region\cite{vernazza:1981, avrett:2008}. The plasmas are partially ionized and the ionization degree of the hydrogen varies from $10^{-4}$ to about $1$. The C7 model based on the results of solving optically thick non local thermodynamic equilibrium (non-LTE) radiative transfer equations and the observational datas is presented in Figure~\ref{fig2}. It shows the variations of the plasma parameters with height in the quiet sun region \cite{avrett:2008}. As the time and space resolutions of the solar telescopes are getting higher and higher, more and more small scale solar activities have been observed, e.g., chromosphere jets/spicules/surges, Ellerman bombs, microflares, UV bursts and so on. On the other hand, the low solar atmosphere is not isolated, it connects with the inner region of the sun and the solar corona. The magnetic flux emerging from the inner region of the sun exists all the time, and reconnection events appear between the emerged magnetic fields with opposite directions, or when the emerged magnetic fields interact with the back ground magnetic fields. Heat and energetic plasmas can conduct and flow along the field lines through the lower solar atmosphere to the corona respectively. The universal reconnection events which connect the low solar atmosphere and the solar corona might be important to corona heating and might contribute to the solar wind. The recent high resolution observations from Parker Solar Probe (PSP) have revealed the fine structure of corona rays, they have suggested that the additional rays drifting toward the southern polar regions are caused by the localized activities in the lower solar atmosphere\cite{poirier:2020}.

\begin{figure}[!ht]
 \centering\includegraphics[width=2.2in]{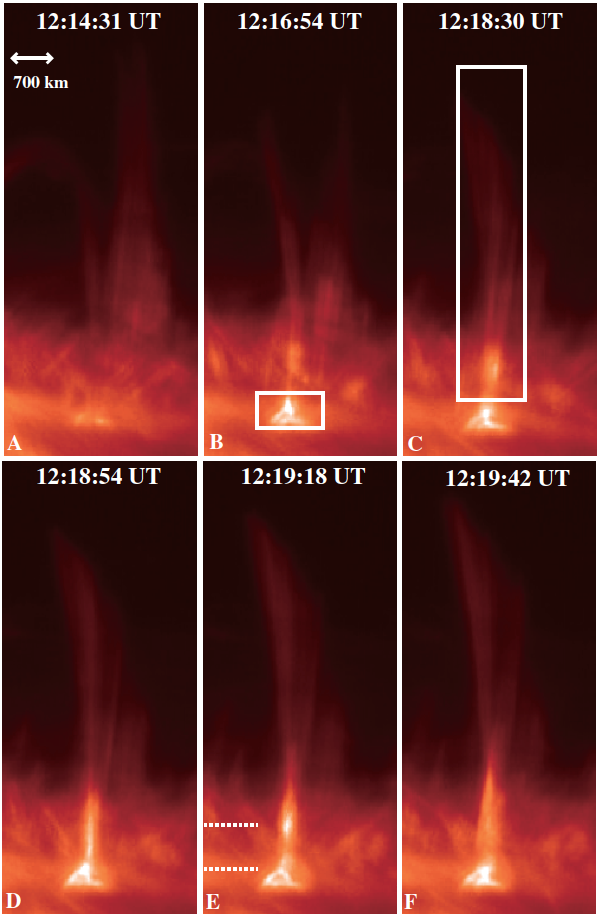}
 \caption{The time evolution of a chromospheric anemone jet observed by SOT/Hinode. The bright blob in this jet is clearly identified. Image reproduced with permission from Singh \textit{et al}. (2012), copyright by AAS}
 \label{fig3}
\end{figure}

The width of the chromospheric jets/spicules/surges is about $0.1-0.4$~Mm, the typical length is about $1-10$~Mm, and the speed is usually about $5-100$\,km/s\cite{de:2011, morton:2012, singh:2012, tian:2014}. The chromospheric jets/spicules/surges material can possibly be heated above $\sim 0.1$~MK\cite{de:2011, morton:2012, tian:2014}. The chromospheric jets/spicules/surges are usually considered to be triggered by magnetic reconnection between the emerged magnetic flux and the surrounding background magnetic fields, the bidirectional flows and the brightening at the foot points suggest that reconnection appears there\cite{tian:2018a}, and the magnetic fields with opposite directions approaching each other can also be observed in photosphere and chromosphere\cite{robustini:2018}. The recent observations by using high resolution data from Goode Solar Telescope (GST) at the Big Bear Solar Observatory and the Solar Dynamic Observatory (SDO) further suggest that magnetic reconnection in partially ionized lower solar atmosphere causes the formations of the spicules and heat the solar corona \cite{samanta:2019}. Figure~\ref{fig3} shows the time evolution of a chromospheric anemone jet observed by SOT/Hinode on January 14, 2007\cite{singh:2012}.  

An Ellerman Bomb (EB)\cite{ellerman:1917, mcmath:1960, georgoulis:2002, rutten:2013} is a transient and prominent enhancement in brightness of the far wings of spectral lines, in particular the Balmer H${\alpha}$ line. They are usually observed in solar active regions and take place exclusively in the region of the new emerging flux\cite{mcmath:1960, rutten:2013, pariat:2004}. They manifest a characteristic elongated shape when shown in high resolution images. The magnetic field topology of a magnetic reconnection process in EBs is usually suggested as the U type called ``bald patchs,'' which is formed due to the dips of serpentine field lines from flux emergence processes\cite{georgoulis:2002, pariat:2004, zhao:2017}. The temperature of an EB is lower than $10^4$~K according to a semi-empirical model of a variety of combinations of the spectral lines\cite{fang:2006, hong:2014, nelson:2015}. 

\begin{figure}[!ht]
 \centerline{\includegraphics[width=3.5in]{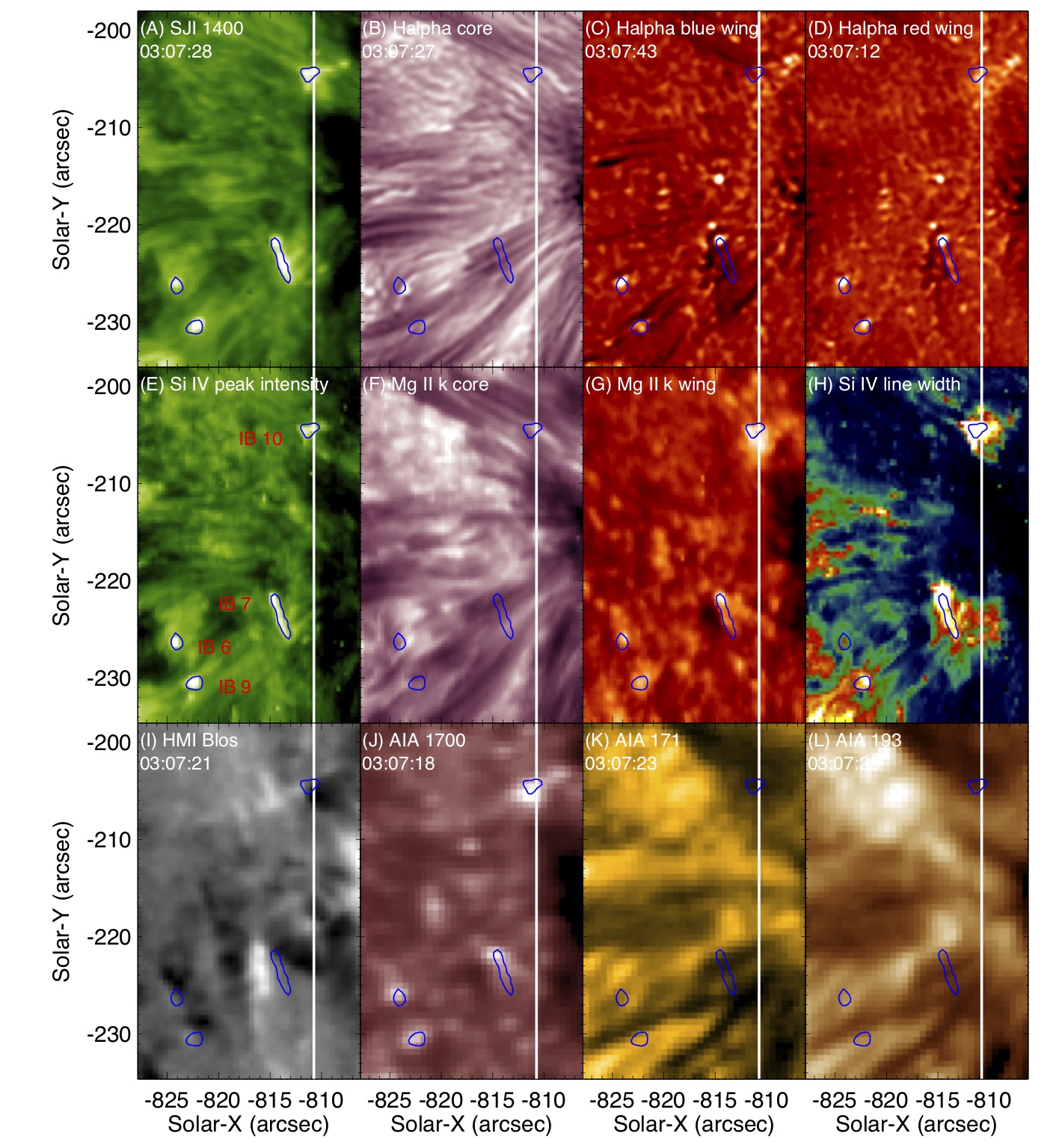}}
  \centerline{\includegraphics[width=3.0in]{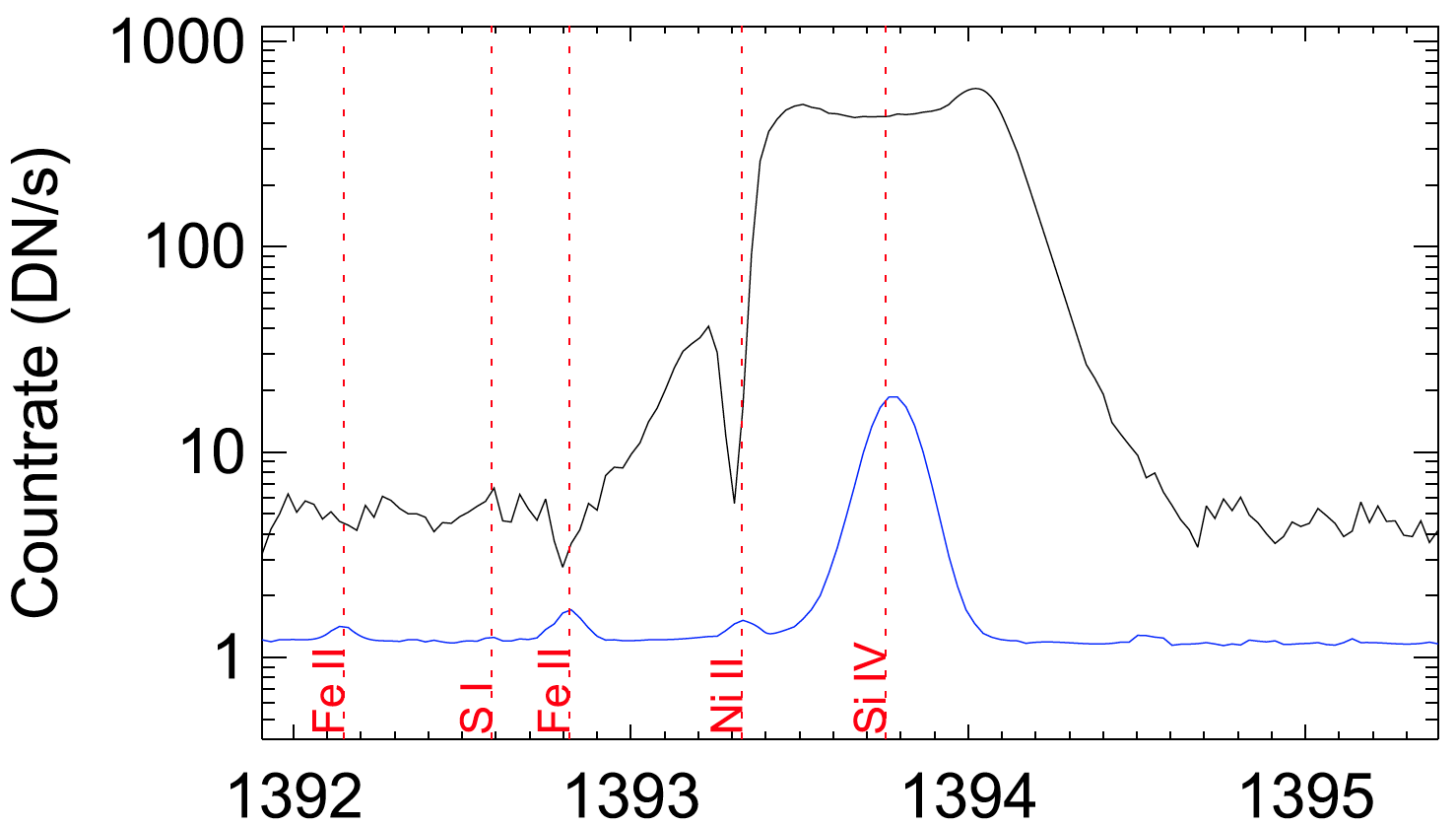}}
 \caption{(A)-(D) IRIS/SJI 1400 \r{A} image, and Chinese New Vacuum Solar Telescope (NVST) H${\alpha}$ core and wing (-1 and +1 \r{A}) images taken around 03:05:38 UT on 2015 May 2. (E)-(H) Images of the Si IV 1393.755 \r{A} intensity, Mg II k core and wing (sum of -1.33 and +1.33 \r{A}), and Si IV 1393.755 \r{A} line width. (I)-(L) \emph{Solar Dynamics Observatory} (\emph{SDO}/HMI) line of sight magneto gram, and \emph{SDO}/AIA images taken around 03:05:38 UT. The white line in these two dimensional images indicates the slit location at the corresponding time. The bottom panel shows the typical IRIS line profile (black line) of a UV burst relating with an EB. The blue line profile represents the reference spectra obtained in a quiet plage region. Image and caption reproduced with permission from Tian \textit{et al}. (2016), copyright by AAS.}
 \label{fig4}
\end{figure}

Recently, the IRIS satellite has discovered many small scale brightenings which are called UV bursts\cite{peter:2014, young:2018, vissers:2015, grubecka:2016, tian:2016}. The intensity of UV burst is 100-1000 times higher than the median background\cite{peter:2014}. They share some characteristics with traditional EBs (e.g., similar lifetimes and sizes), and they are also considered to be formed in magnetic reconnection processes. UV bursts can have the U type or the fan-spine magnetic topology\cite{chen:2019a, chitta:2017}. The strong emissions in Si IV lines require a temperature of at least about $2\times10^4$~K around the solar temperature minimum region and about $8\times10^4$~K in the transition region\cite{rutten:2016}. Though UV bursts have a signature in the UV continua at 1600 and 1700 observed by the SDO Atmospheric Imaging Assembly (AIA), but remain invisible in its He II and higher-temperature coronal channels. The presence of Fe II and Ni II absorption lines and the O IV forbidden lines indicate that the high density and cool plasmas \cite{tian:2016, tian:2018b, judge:2015} also exist in the reconnection region for generating UV bursts.  The spectral profiles of  Si IV, C II and Mg II h\&k emission lines in UV burst are significantly enhanced and broadened\cite{peter:2014, tian:2016}, the line width of the Si IV spectral profile is usually about 100 km\,s$^{-1}$ away from line center. The commissioning observations by using the CHROMIS instruments at the Swedish 1-m Solar Telescope (SST) uncover a whole new level of fine structures in UV bursts, which are highly dynamic blob-like substructure evolving on the timescale of seconds \cite{rouppe:2017}. The non-LTE inversions of low-atmosphere reconnection based on the SST and IRIS observations also shown the blob-like substructures\cite{vissers:2019}. These blob-like fine structures are suggested as one of the mechanisms to cause the non-Gaussian broadened spectral profiles. Figure~\ref{fig4} shows the intensity distributions of three UV bursts relating with EBs in different wavelengths\cite{tian:2016}. We should also mention that the UV bursts with narrow Si IV line profiles are also observed \cite{hou:2016}. The recent observations show that some low solar atmosphere reconnection events also have UV brightenings, but they probably occur at higher locations than where UV bursts usually appear\cite{guglie:2018, guglie:2019, lid:2018}, which explains the observed coronal counterparts in this kind of reconnection events.       

The joint observations by using IRIS satellite and ground-based telescopes (e.g., SST, NVST) show that some UV bursts have strong emissions both in the wings of H${\alpha}$ and the IRIS Si IV passbands\cite{vissers:2015, tian:2016}. The observations showed that the occurrence of cospatial and cotemporal EBs and UV bursts is between $10\%$ and $20\%$ \cite{vissers:2015, tian:2016, chen:2019b}, which suggests that the UV bursts  and their associated EBs are caused or modulated by a common physical process. However, the observed H${\alpha}$ emission cannot be reproduced using non-local thermal equilibrium (non-LTE) semi-empirical modeling if the temperature is above $10^4$~K\cite{fang:2017}. Therefore, the H${\alpha}$ emissions and Si IV emissions in these events should come from the plasmas with different temperatures. Recently, 161 EBs has been identified by using data from GST, and ~20 of them reveal signatures of UV bursts in the IRIS images. The authors found that most of these UV bursts have a tendency to appear at the upper parts of their associated flame-like EBs. 

There is another kind of reconnection events, which are considered to be formed in the upper chromosphere or transition region (Transition region explosive events)\cite{brueckner:1983, huang:2014}. They have extended strong wings (50-250 km\,s$^{-1}$) in the transition region spectra indicating the existence of ``bi-directional'' flows. The small scale current sheet like structures between loops and filaments have also been observed through different wavelengths\cite{yang:2015, xue:2016, xue:2018, huang:2018}. The measured reconnection inflow and outflow speeds and plasma density indicate that parts of these current sheets are located below the transition region. Numerous small scale reconnection events with different magnetic topologies and triggering mechanisms exist in the low solar atmosphere\cite{huang:2019}. However, we still know little about the diffusion regions in the reconnection process from observations because of the limited space and time resolutions. The highest space resolution of the existing solar telescopes is about dozens of kilometers, which is still much larger than the ion-neutral collision mean free path ($ \lesssim10$\,m) and the ion inertial length ($ \lesssim1$\,m) in the low solar atmosphere. The measurement of magnetic field especially the horizontal one is also very difficult, the best resolution for the photosphere magnetic field is only about $10$~G along the line of sight. Therefore, we need theoretical and numerical simulations, as well as improvements of magnetic field measurements, to study the physical mechanisms of magnetic reconnection in the low solar atmosphere and how the reconnection events are triggered and formed.

 \subsection{Theories and numerical simulations}
\begin{figure}[!ht]
 \centerline{\includegraphics[width=2.5in]{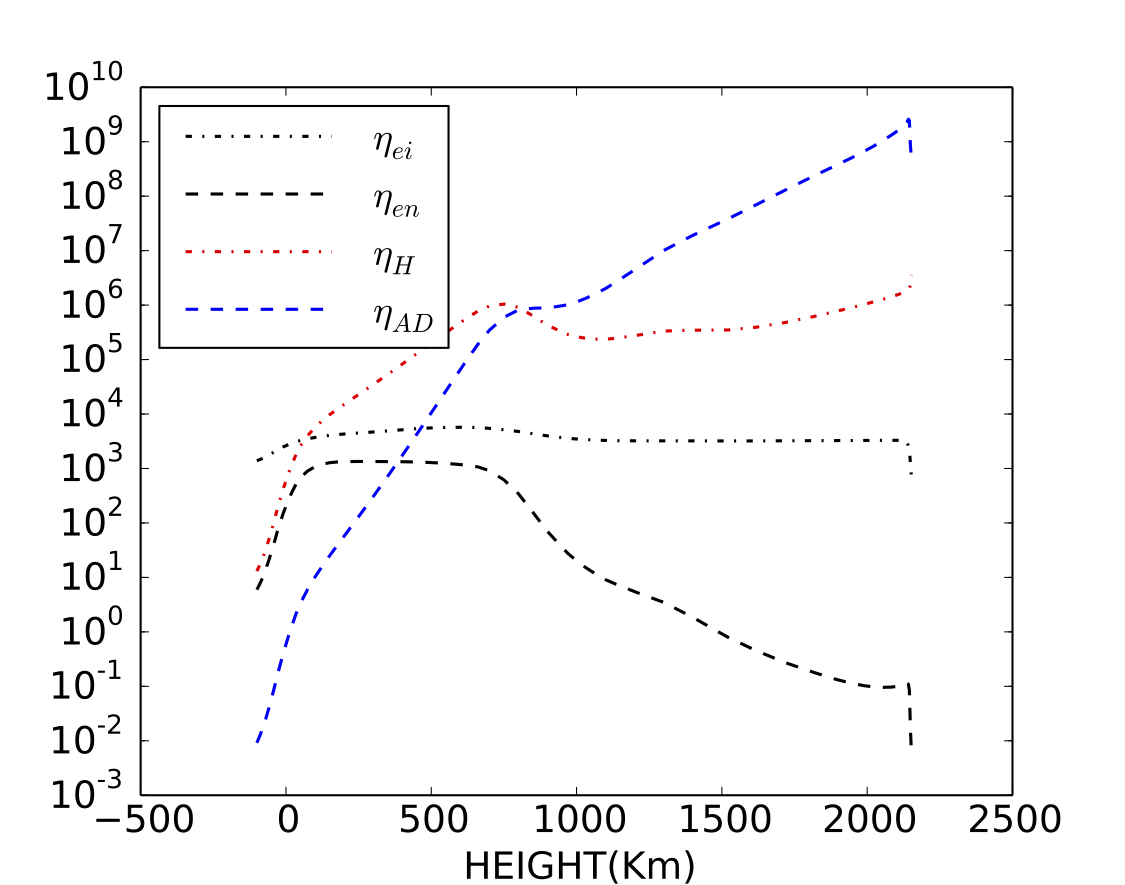}
             \includegraphics[width=2.5in]{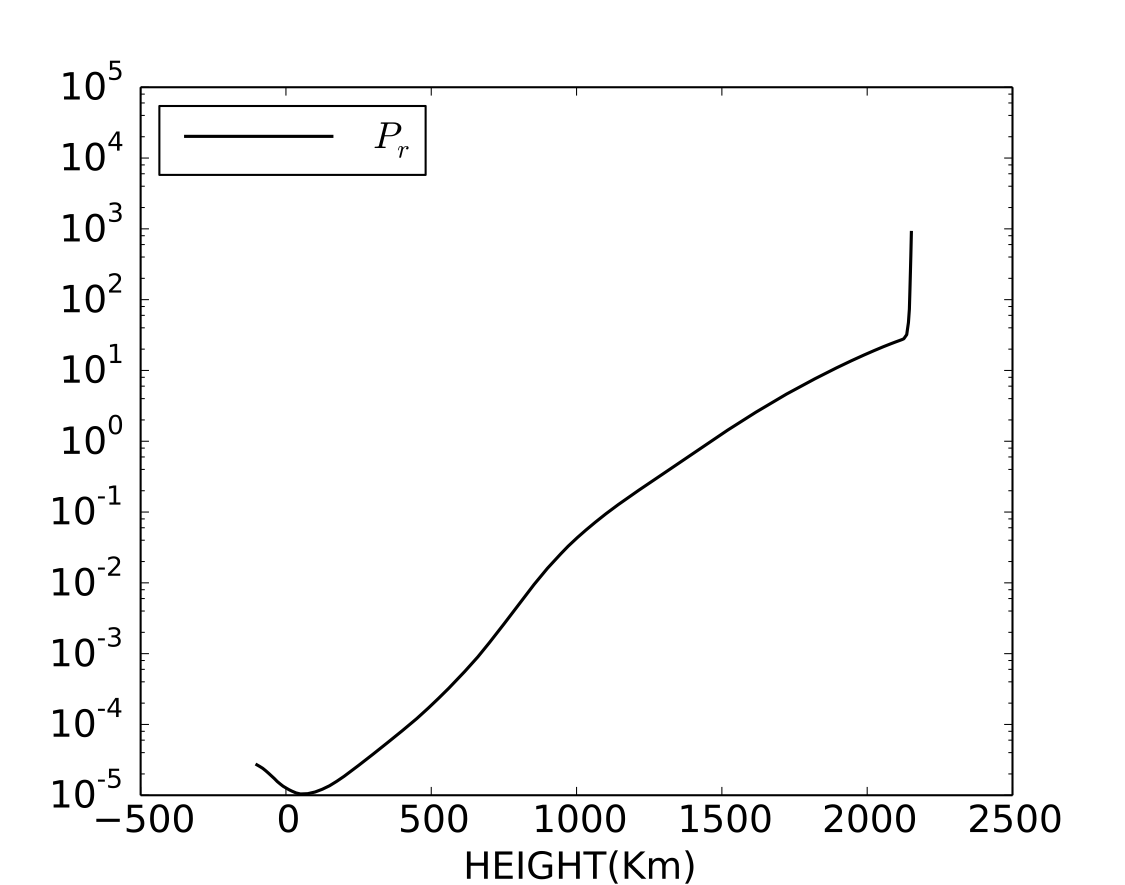}}
 \caption{Panel (a) shows the variations of different diffusivities with height, panel (b) shows the variation of the magnetic Prandtl number $P_r$ with height. The plasma parameters are based on the C7 Model and the magnetic field is $100$~G. }
 \label{fig5}
\end{figure}

As mentioned in the first paragraph of the last subsection, the plasmas are highly stratified and the plasma density is much higher in the low solar atmosphere than in the solar corona. The plasma temperature in these deep solar atmospheric layers is only about several thousand Kelvin and partially ionized. Therefore, several physical mechanisms which are not important or do not exist in the solar corona can become very important when we study reconnection mechanisms in the low solar atmosphere. These important physical mechanisms include ambipolar diffusion, non-equilibrium ionization, radiative cooling, magnetic diffusion contributed by ion-neutral collisions, viscosity, gravity, charge-exchange and so on. The different diffusion coefficients which are contributed by electron-neutral collisions, electron-ion collisions, ambipolar diffusion and Hall effect are plotted in Figure~\ref{fig5}(a), where $\eta=\eta_{ei}+\eta_{en}$, $\eta_{ei}=m_e \nu_{ei}/(e^2 n_e \mu_0)$, $\eta_{en}=m_e\nu_{en}/(e^2n_e\mu_0)$, and the Hall effect coefficient $\eta_H=\lvert \mathbf{B} \rvert/(en_e\mu_0)$. The strength of magnetic field is assumed as 100 G  and the C7 atmosphere model \cite{avrett:2008} is applied to plot this figure. It shows that the magnetic diffusion $\eta$ is mainly contributed by collisions between electrons and ions even in the photosphere. One can also find that the diffusivity contributed by Hall effect is significant above the solar surface, $\eta_H$ is about two orders of magnitude larger than $\eta$ above the solar temperature minimum region (TMR). Ambipolar diffusivity $\eta_{AD}$ is smaller than the magnetic diffusivity $\eta$ in the photosphere. However, $\eta_{AD}$ significantly increases with height and it becomes much larger than both $\eta_H$ and $\eta$ above the middle chromosphere. In the up chromosphere, $\eta_{AD}$ is about five orders of magnitude higher than $\eta$. One should note that both $\eta_H$ and $\eta_{AD}$ depend on the strength of magnetic fields, the stronger magnetic fields will make $\eta_H$ and $\eta_{AD}$ to be larger. In the previous paper\cite{khomenko:2012}, the thin flux rope model with different strengths of magnetic fields at different heights has been applied to calculate the distributions of these coefficients with height \cite{khomenko:2012}, the results in this paper are similar as shown in Figure~\ref{fig5}(a). However, one should keep in mind that the C7 atmosphere model is for the quiet sun region, the plasma parameters (e.g., temperature and ionization degree) might change dramatically with time in the reconnection region. The recent numerical simulations \cite{ni:2018a, ni:2018b, ni:2018c} show that the neutrals decrease and the ionized plasmas strongly increase with time in the reconnection region when the reconnection magnetic field is strong enough, which makes both $\eta_{AD}$ and $\eta_{H}$ to be much smaller than $\eta$ around the temperature minimum region. The viscosities which are contributed by the collisions between each particle might also affect the reconnection process.  The neutral and ion viscosity coefficients are given by $\xi_n=n_nk_BT_n/\nu_{nn}$ and $\xi_i=n_ik_BT_i/\nu_{ii}$, respectively. The relevant collision frequencies $\nu_{nn}$ and $\nu_{ii}$ are given by $\nu_{nn}=n_n\sigma_{nn}\sqrt{\frac{16k_BT_n}{\pi m_n}}$ and $\nu_{ii}=\frac{4}{3}n_i\sigma_{ii}\sqrt{\frac{2k_B T_i}{\pi m_i}}$. The variation of magnetic Prandtl number $P_r$ with height is calculated and plotted in Figure~\ref{fig5}(b). $P_r$ is defined by the ratio of viscous to resistive diffusivity,  $P_r=\frac{\xi}{\eta}\frac{1}{(n_im_i+n_nm_n)}$, where $\xi=\xi_i+\xi_n$. Figure~\ref{fig5}(b) demonstrates that the effect of the viscosity is much less important than the magnetic diffusion below the middle chromosphere. However, the magnetic Prandtl number $P_r$ is much larger than one in the upper chromosphere, which means that the viscosity might play some role in the reconnection process in this region. In Figure~\ref{fig5}(b), we assume the ion and neutral temperatures are the same as shown in C7 model, and the cross sections $\sigma_{nn}$ and $\sigma_{ii}$ are assumed the same as those in the previous paper\cite{leake:2013}. 

Comparing with the solar corona, there are much fewer works about theories and numerical simulations of magnetic reconnection in the low solar atmosphere. Some of the numerical works focus on more about the triggering and the formation process of different kinds of reconnection events and compare the simulation results with observations, while the others study magnetic reconnection mechanisms in the diffusion region.  
 
\begin{figure}[!ht]
 \centering\includegraphics[width=5.0in]{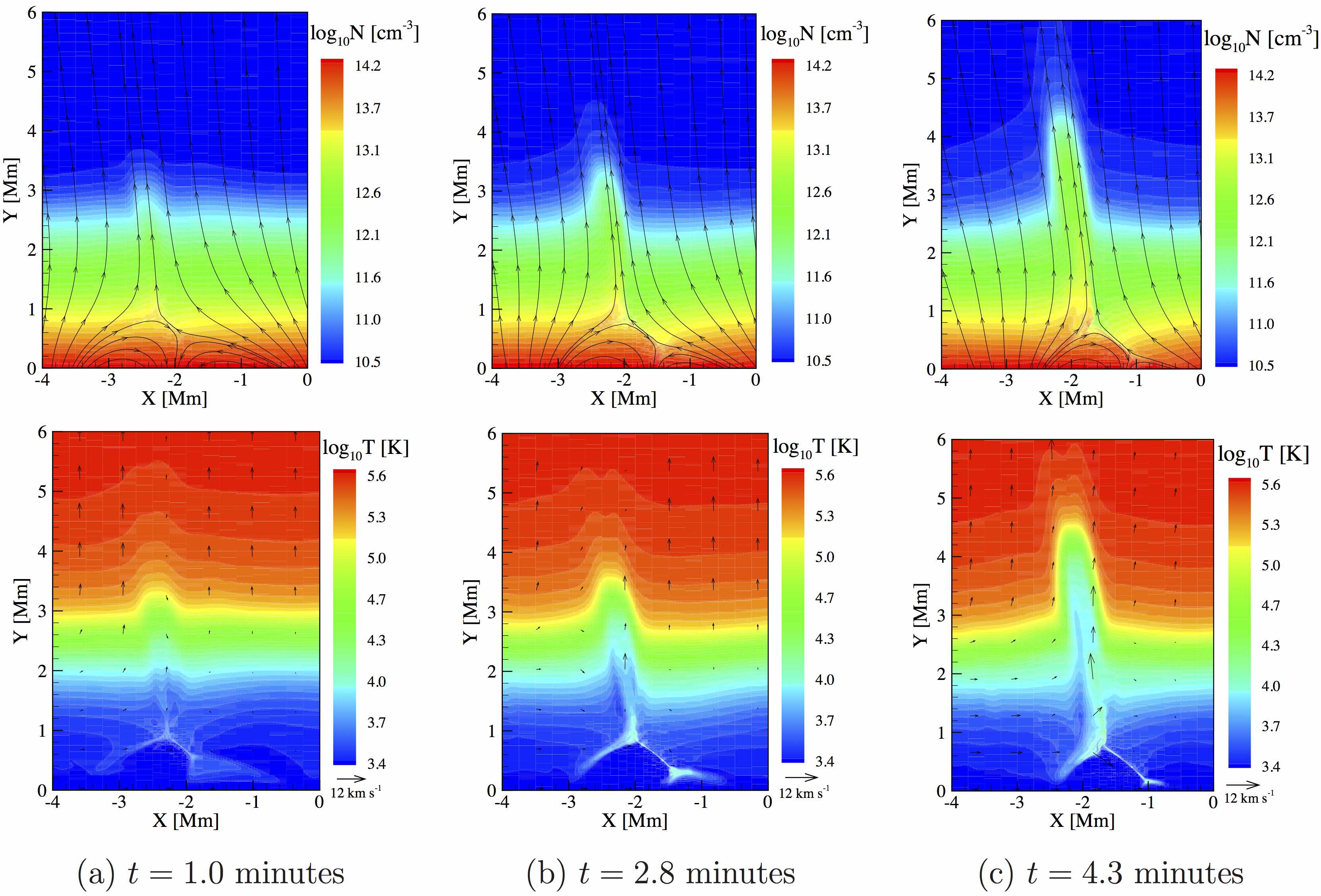}
 \caption{Calculated distributions of number density $N$ and temperature $T$ at $t=1.0$ minute, $t = 2.8$ minutes, and $t=4.3$ minutes, where streamlines in the top row denote magnetic field lines and black arrows in the bottom row indicate the total velocity, respectively. Image and caption reproduced with permission from Yang \textit{et al}. (2013), copyright by AAS.}
 \label{fig6}
\end{figure}

The formation of chromospheric jets/spicules/surges has been studied in 2D and 3D simulations with the one-fluid magnetohydrodynamic (MHD) model\cite{nishizuka:2008, ding:2011, takasao:2013, yang:2013, nobrega:2016, nakamura:2012, mac:2015}. Most of these simulations are based on the scenario of the standard jet model proposed by Yokayama \& Shibata \cite{yokoyama:1995}, in which magnetic reconnection between the emerging magnetic flux and pre-existing magnetic fields triggers the formation of jets. Different from the coronal jets, magnetic reconnection for triggering chromosphere jets happens in the chromosphere, which makes the length scale and jet speed smaller, the lifetime shorter and the plasma temperature of the jet lower than the coronal jet \cite{singh:2011}. Most of the characteristics of the  chromospheric jets/spicules/surges can be reproduced in the present numerical simulations. The jet speed is close to the local Alfv\'en speed, the observed evidences for Alfv\'en waves propagating along the jet are confirmed in the jet simulations\cite{nishizuka:2008}. The plasmoids which are generated in the reconnection process at the foot point of the jet in the simulations \cite{yang:2013, nobrega:2016} are believed to correspond to the downward and upward bright blobs from observations. Most of jets have both cool and hot components, there are several possible mechanisms to result in this phenomenon according to numerical results. The early numerical simulations show that the hot jets are produced due to magnetic reconnection and the cool jets are due to the slingshot effect\cite{yokoyama:1995}. Recently, 2.5 D numerical simulations with a realistic treatment of the radiative transfer and material plasma properties show how the formation of the cool ejection is mediated by a wedge-like structure composed of two shocks in the upward reconnection outflow region\cite{nobrega:2016}. The filament eruption process may also provide the material of the cool jet \cite{shen:2012, shen:2017, wyper:2018}. Some characteristics from observations can not be simulated in 2D, e.g., the multi-thread structures of jets, the helical or whip-like motion of the jets and the changes of the location of the jet foot point. A 3D reconnection model with more complicated magnetic field structures has to be applied to study these characteristics. Figure~\ref{fig6} shows the simulation results of a chromosphere jet triggered by magnetic reconnection\cite{yang:2013} in 2D simulations. 

EBs have also been numerically studied based on one-fluid magnetohydrodynamic (MHD) model. The partial ionization effect with the time dependent ionization degree and simple radiative cooling have been included in the 2D one-fluid MHD simulations to study magnetic reconnection in the low solar atmosphere\cite{chen:2001}, the numerical results indicate that EBs can be formed in a magnetic reconnection process at different heights from the photosphere to the middle chromosphere\cite{chen:2001}. The results also show that the time-dependent ionization degree and radiative cooling have strong effects on magnetic reconnection at around the middle chromosphere\cite{chen:2001}. Based on the simulation results, the spectral profiles of EBs and micro flares are calculated with the non-LTE radiative transfer theory and compared with observations\cite{xu:2011}. It is found that the typical features of the two phenomena can be qualitatively reproduced\cite{xu:2011}. The Parker instability has been applied to trigger flux emergency\cite{isobe:2007, archontis:2009}, then the multiple emerging loops expand in the atmosphere and interact with each other to lead the magnetic reconnection process in the low solar atmosphere with U type magnetic field topology\cite{isobe:2007, archontis:2009}, the density and temperature increase in the reconnection region are also reasonable to reproduce the H${\alpha}$ wings\cite{isobe:2007, archontis:2009}. Recently, the 3D  MHD simulations further confirm the EB-like brightenings by magnetic reconnection, the intensity images in H${\alpha}$ wings and magnetograms obtained from Fe I 6301 \AA\, are synthesized\cite{danilovic:2017a, danilovic:2017b}. The results show that EB features are caused by reconnection of strong-field patches of opposite polarity in the regions where the surface flows are strong\cite{danilovic:2017a, danilovic:2017b}. 

\begin{figure}[!ht]
 \centerline{\includegraphics[width=2.2in]{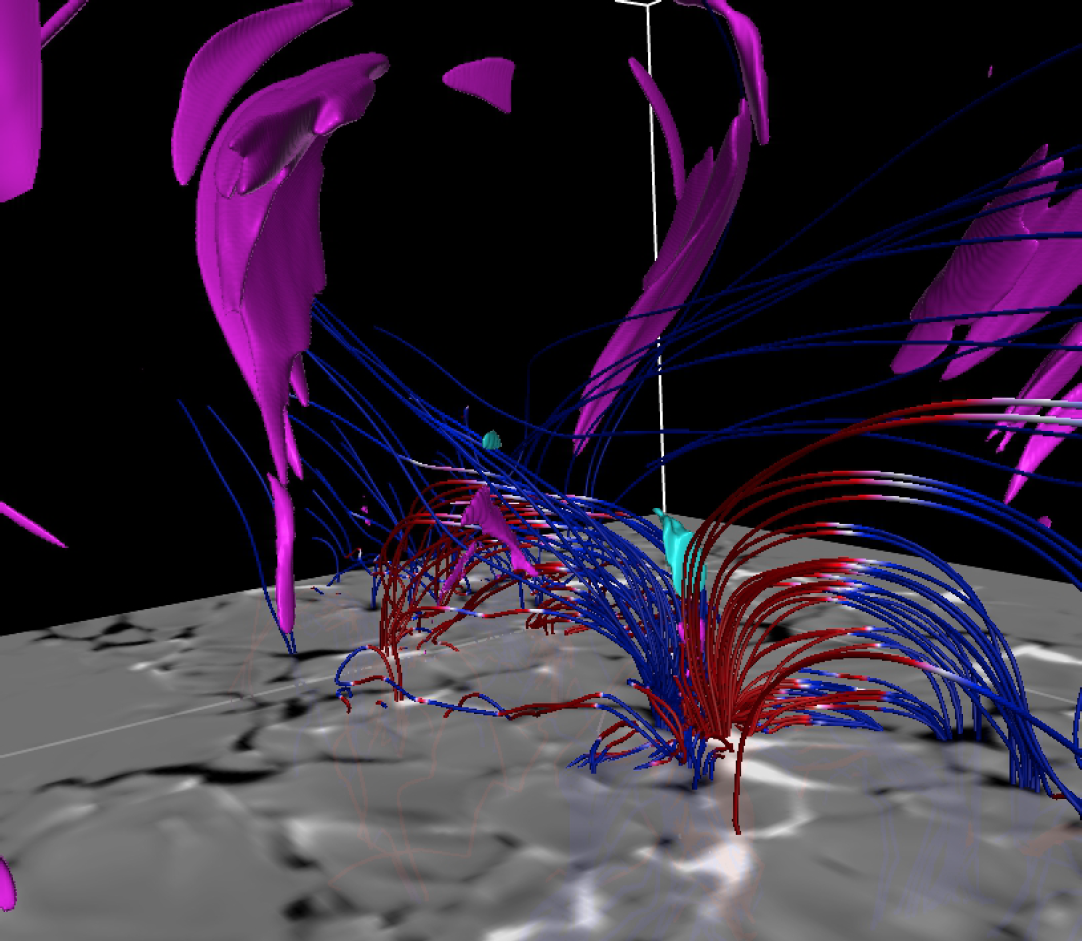}}
  \centerline{\includegraphics[width=2.152in]{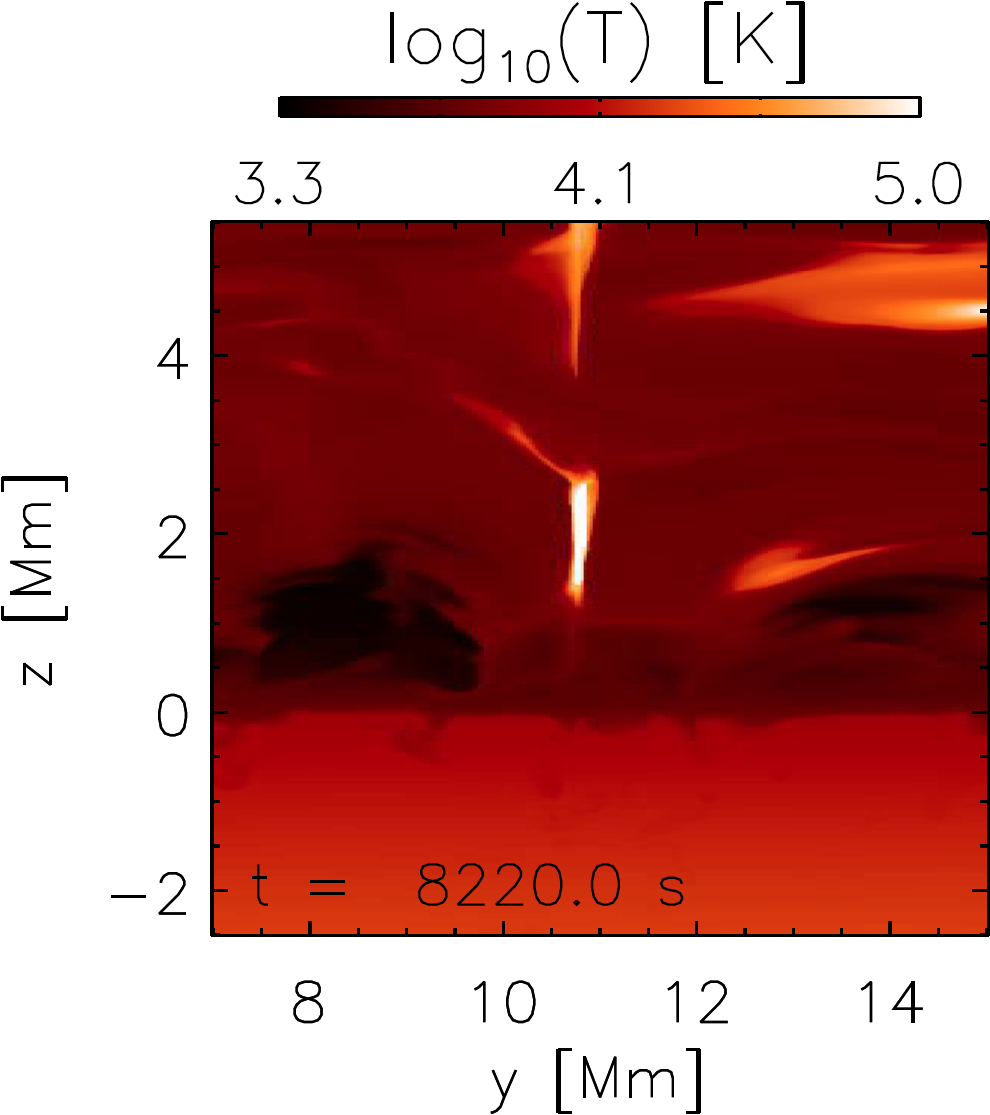}
             \includegraphics[width=1.7in]{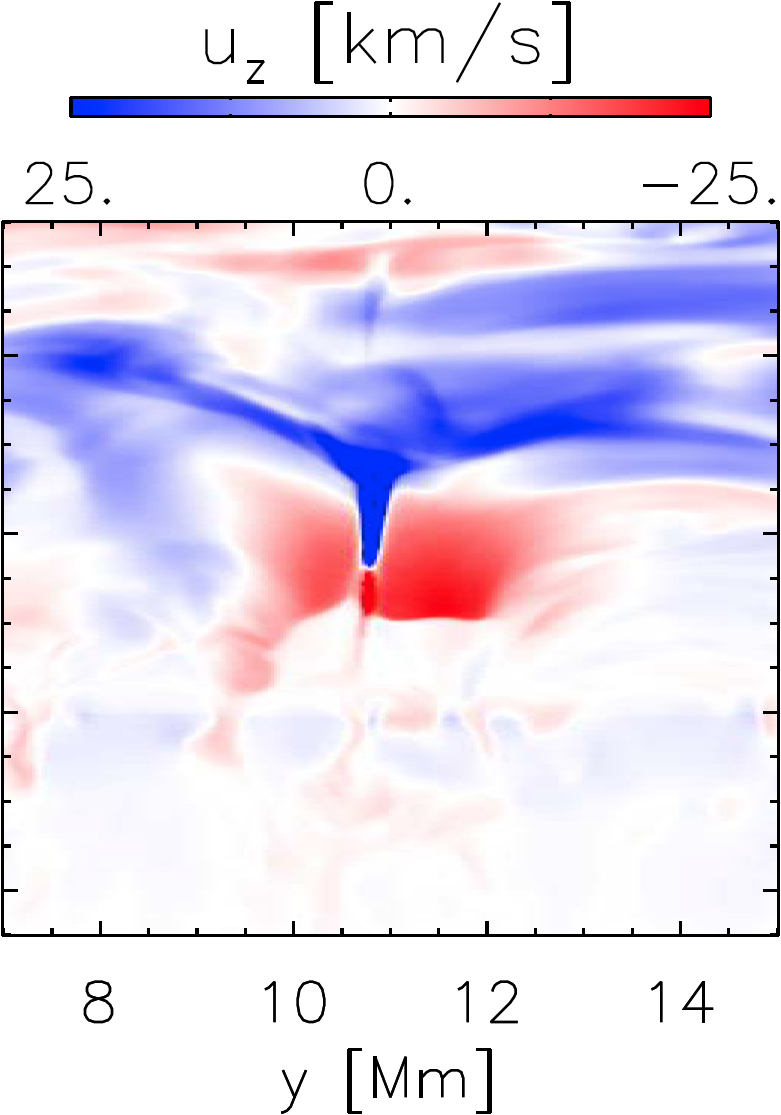}
              \includegraphics[width=1.7in]{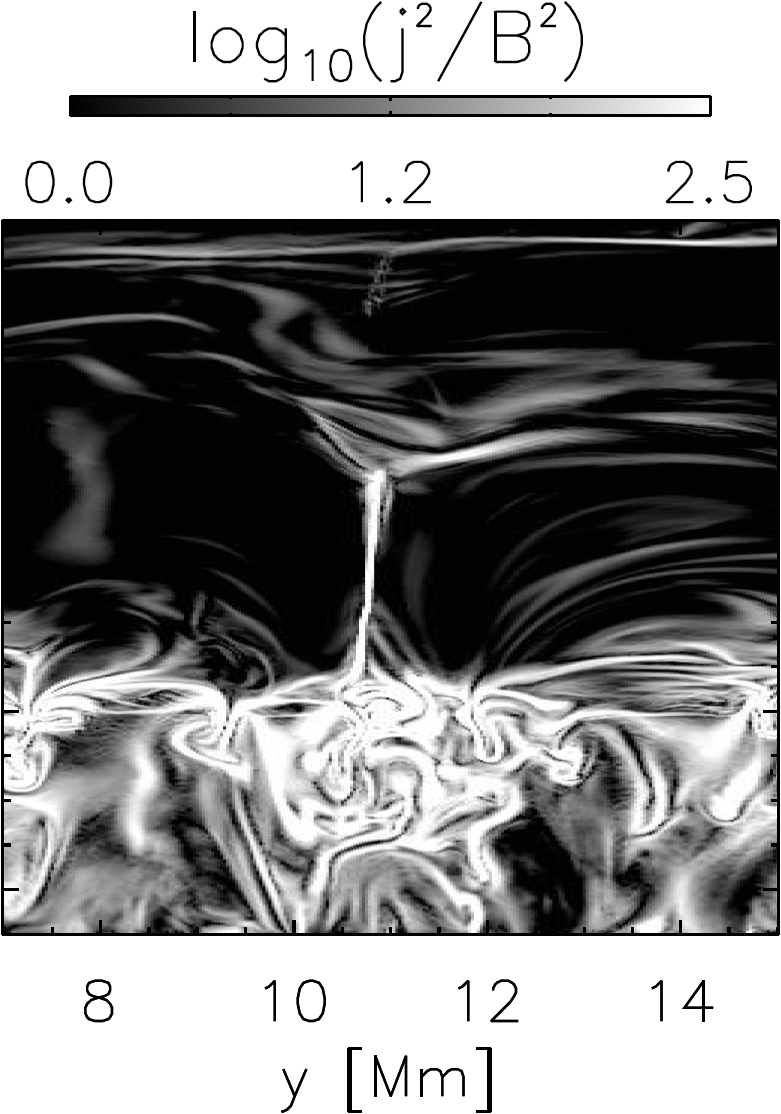}}
 \caption{The top panel shows the up and down flows in and around the current sheet, and magnetic field lines surrounding and above the current sheet. The cyan color represents the isosurfaces of $100$ km/s upward flows and the purple color represent the isosurfaces of $-60$ km/s downward flows. The red and blue lines represent the magnetic field lines with positive and negative $B_z$.  The bottom panels show the two dimensional cuts at $x=10.78$ Mm in the yz plane, of regions in the vicinity of the current sheet at $t=8220.0$~s near the time of maximum intensity of Si IV lines, the distributions of the temperature $T$, vertical velocity $v_z$ and the current strength $j^2/B^2$. Image and caption reproduced with permission from Hansteen \textit{et al}. (2019), copyright by ESO}
 \label{fig7}
\end{figure}

The 3D radiative MHD simulations find that the emerging bipolar magnetic fields can trigger EBs in the photosphere, UV burst in the middle chromosphere and small flares in the upper chromosphere\cite{hansteen:2017}.  The same radiative MHD code is then used to study the UV bursts which connect with EBs\cite{hansteen:2019},  their numerical results show that a long-lasting current sheet that extends over different scale heights through the low solar atmosphere is formed. The part above 1 Mm of such a long current sheet is very hot and the temperature reaches about $10^5$~K, the temperature of the lower part of this current sheet is below $10^4$~K\cite{hansteen:2019}. The synthetic spectra in H${\alpha}$ and Si IV 139.376 nm lines both show characteristics that are typical of the observations of EBs and UV bursts. The simulation results indicate that EBs and UV bursts are occasionally formed at opposite ends of a long current sheet\cite{hansteen:2019}, and this can be one of the possible models to explain the UV bursts which connect with the EBs. Figure~\ref{fig7} shows the simulation results of an UV burst relating with an EB in the 3D simulations\cite{hansteen:2019}.

Most of the numerical simulations which reproduce the reconnection events in the low solar atmosphere are based on the one fluid MHD model. The physical mechanisms resulted by interactions between neutrals and ionized plasmas such as the non-equilibrium ionization, ambipolar diffusion and so on are rarely included in these simulations. Though there are several works about numerical simulations of magnetic flux emergency \cite{leake:2013b, martinez:2017, nobrega:2020}, MHD waves \cite{khomenko:2012} and local dynamo \cite{khomenko:2017}, which have considered the effects of ambipolar diffusion and non-equilibrium ionization. Because of the limited resolution, the artificially assumed magnetic diffusion or numerical diffusion are usually applied to trigger reconnection and heat plasmas. Though the synthetic images and spectra from these numerical simulations are very similar to the observational results, it is still very important and necessary to find out what really happens in the reconnection region and what are the real mechanisms to lead fast reconnection and heat the plasmas.   

Magnetic reconnection in the low solar atmosphere has been analytically studied to explain Type II white-light flares\cite{li:1997}. The derived lifetime, the total energy release based on the analytical model and the plasma parameters in this region agree well with the observations\cite{li:1997}. However, the  collision between the electrons and neutrals is the only considered plasma-neutral interaction effect, and it only contributes a little bit to magnetic diffusion in the low solar atmosphere as show in Figure~\ref{fig5}(a).

\begin{figure}[!ht]
    \centerline{\includegraphics[width=2.2in]{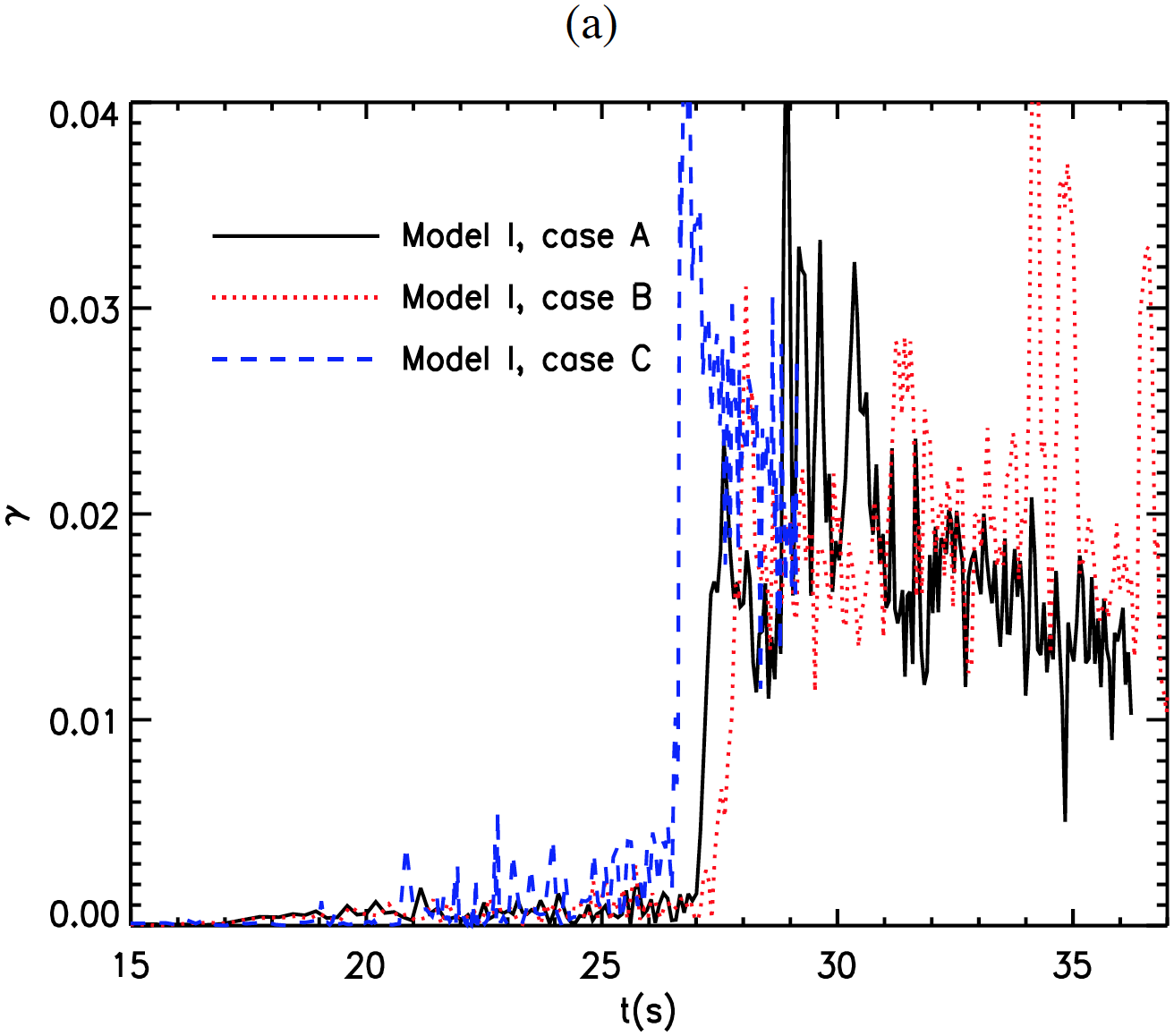}
                       \includegraphics[width=2.2in]{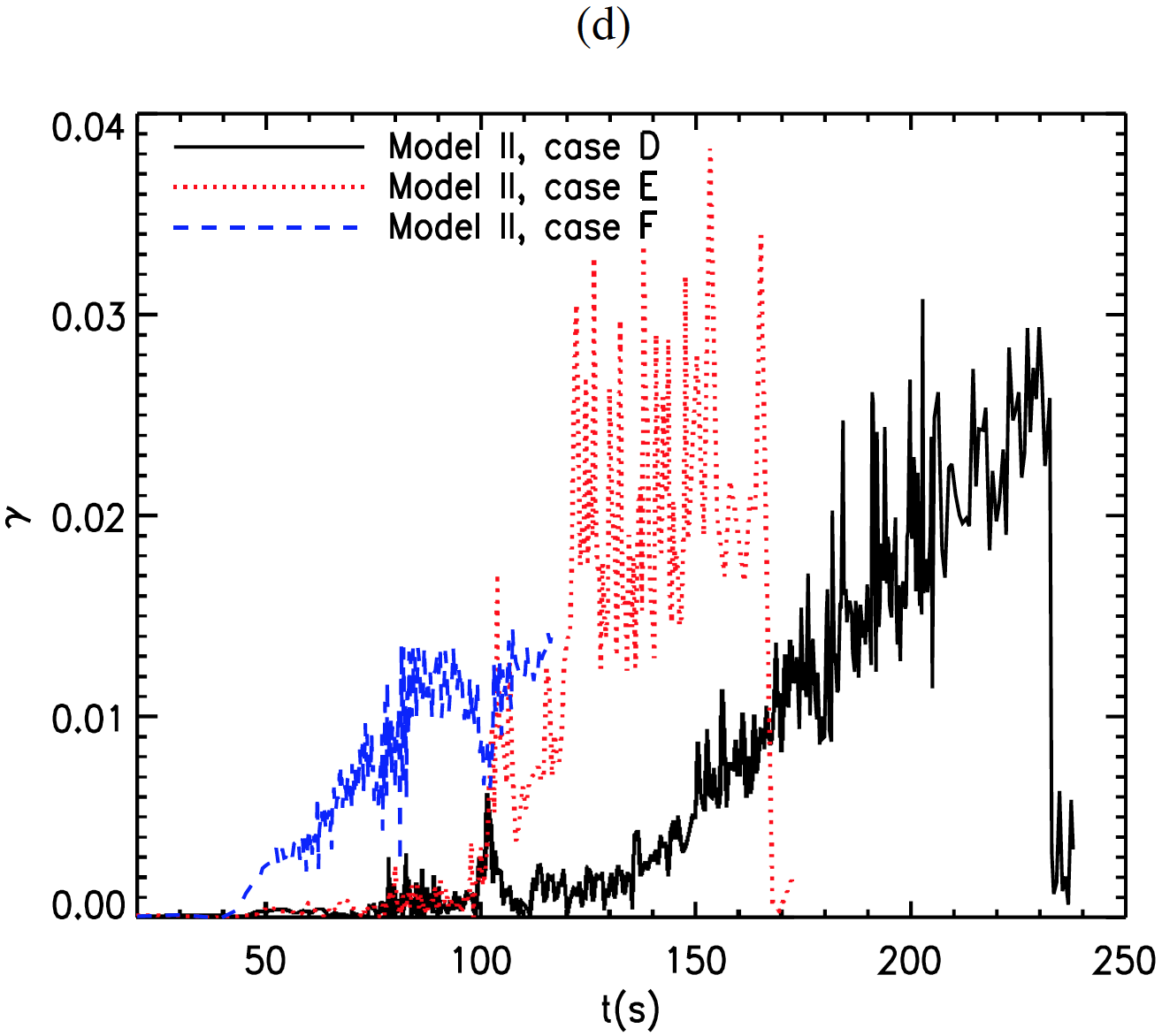}}
    \centerline{\includegraphics[width=2.2in]{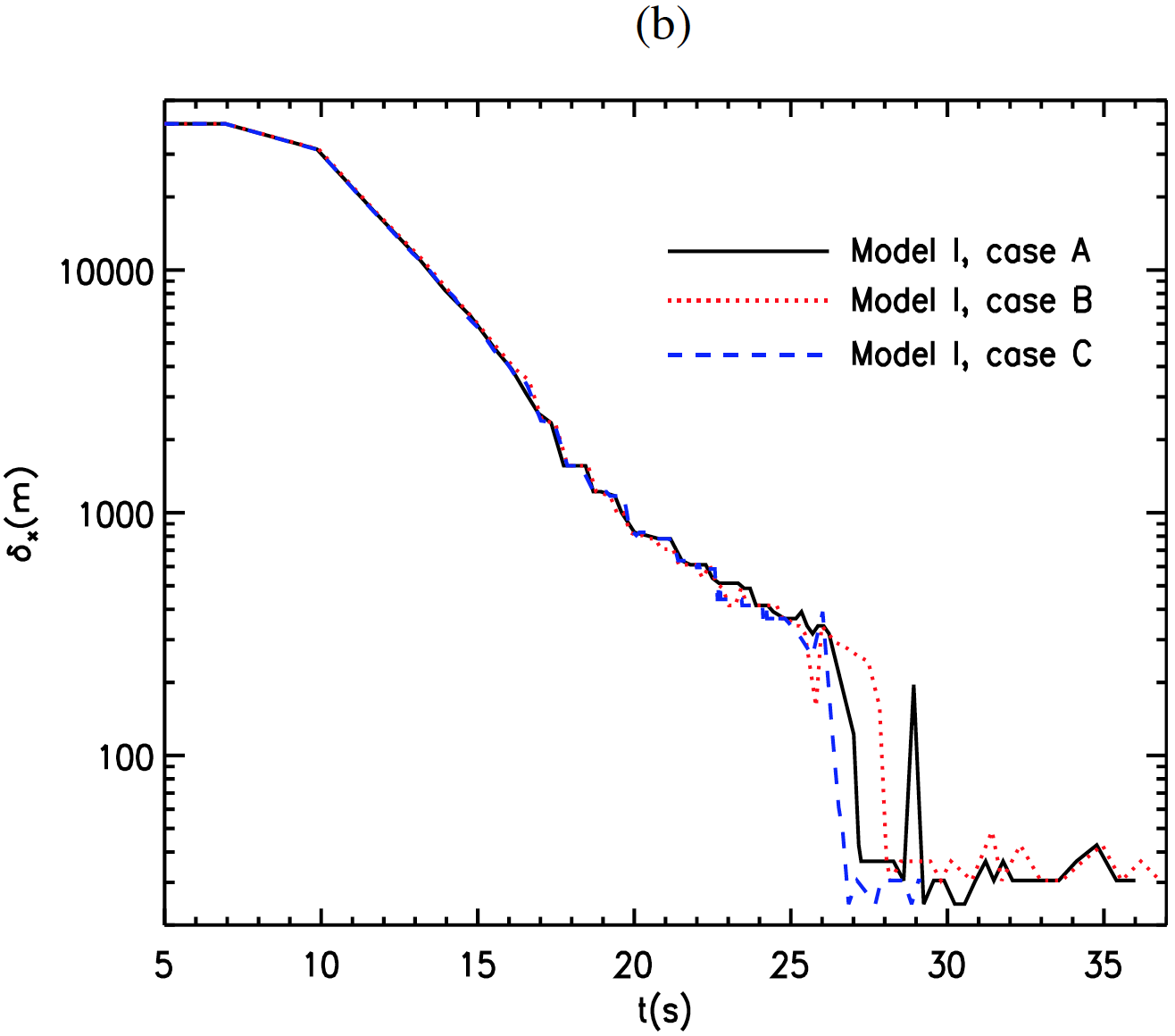}
                       \includegraphics[width=2.2in]{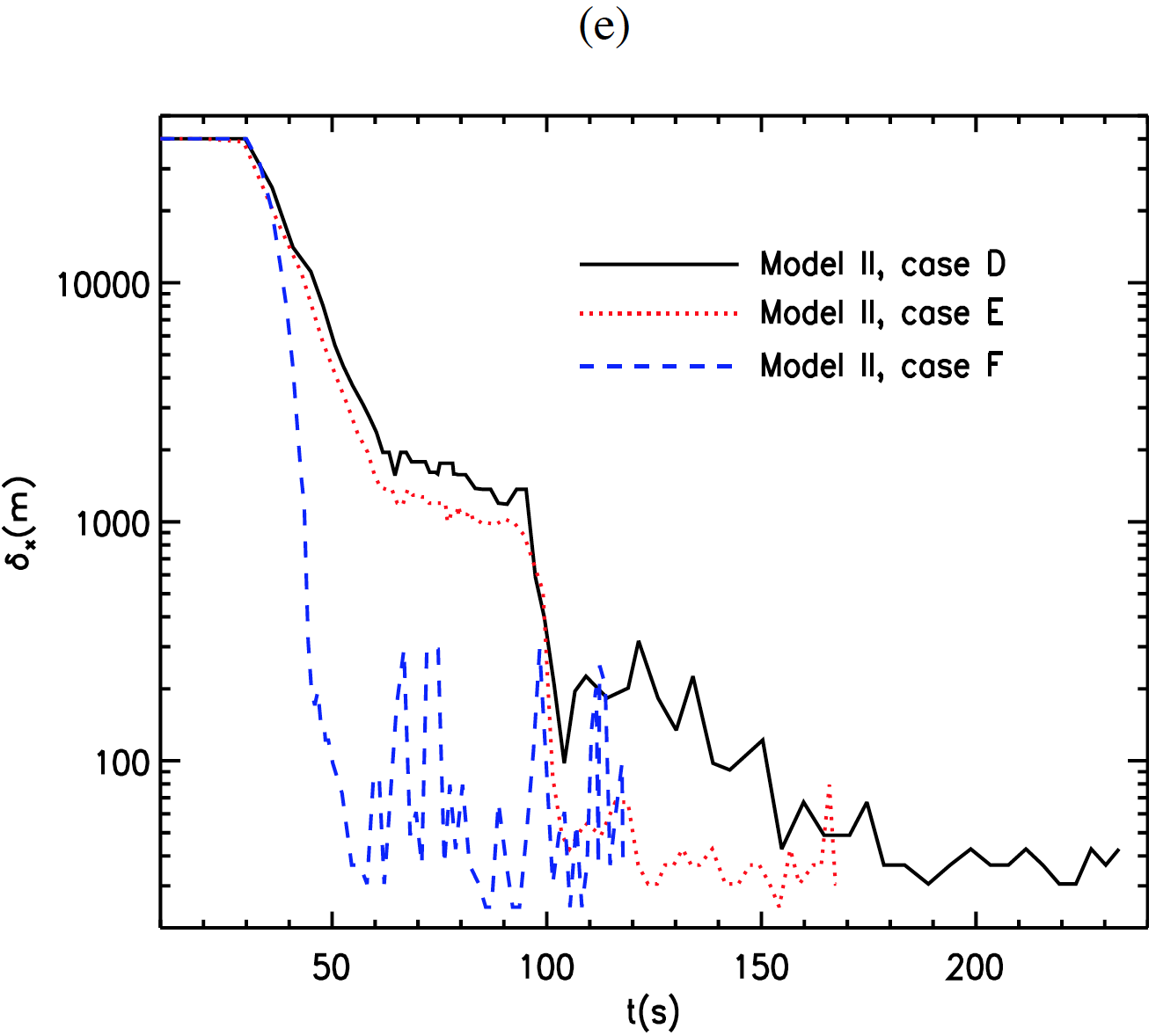}}
     \centerline{\includegraphics[width=2.2in]{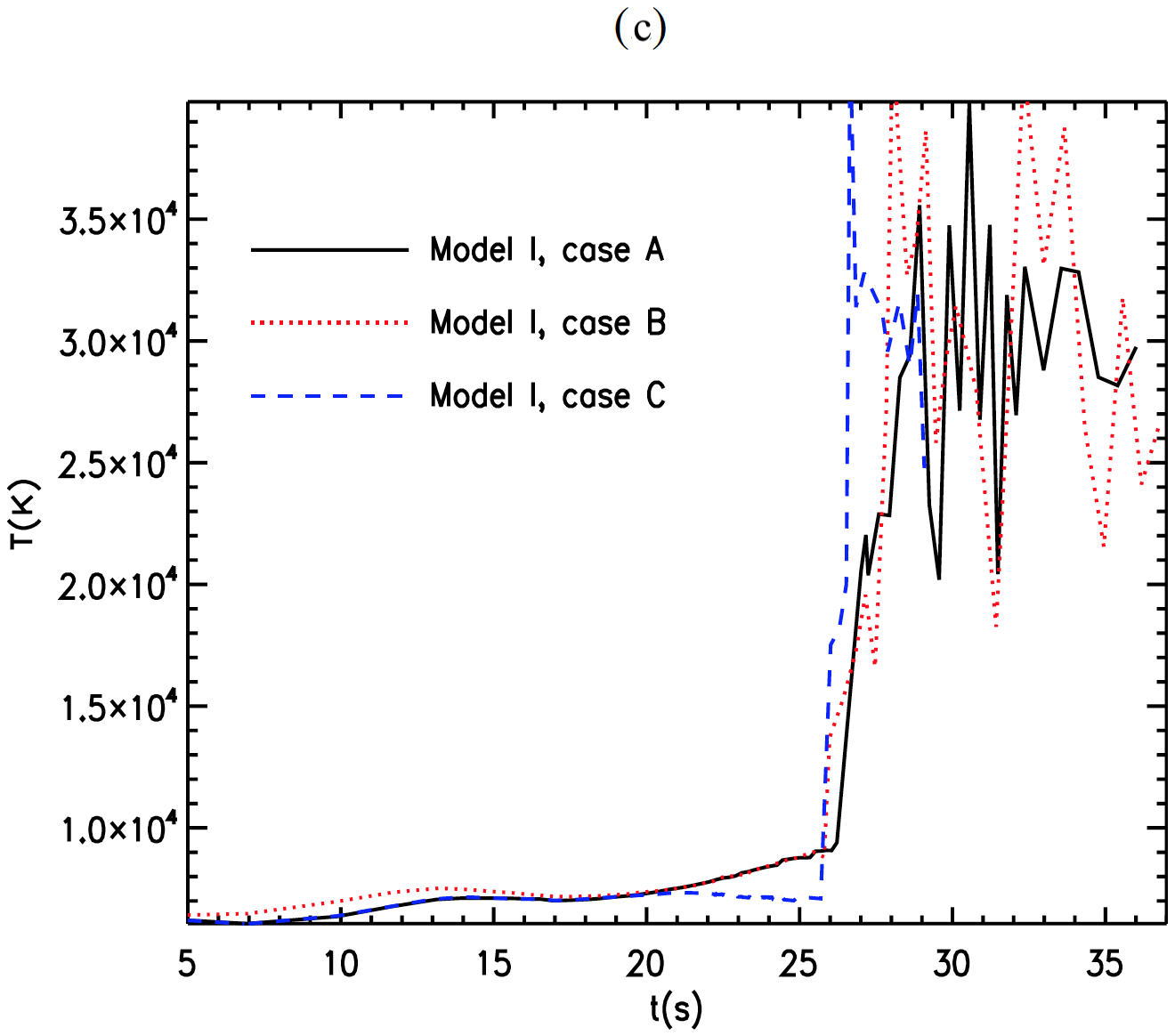}
                       \includegraphics[width=2.2in]{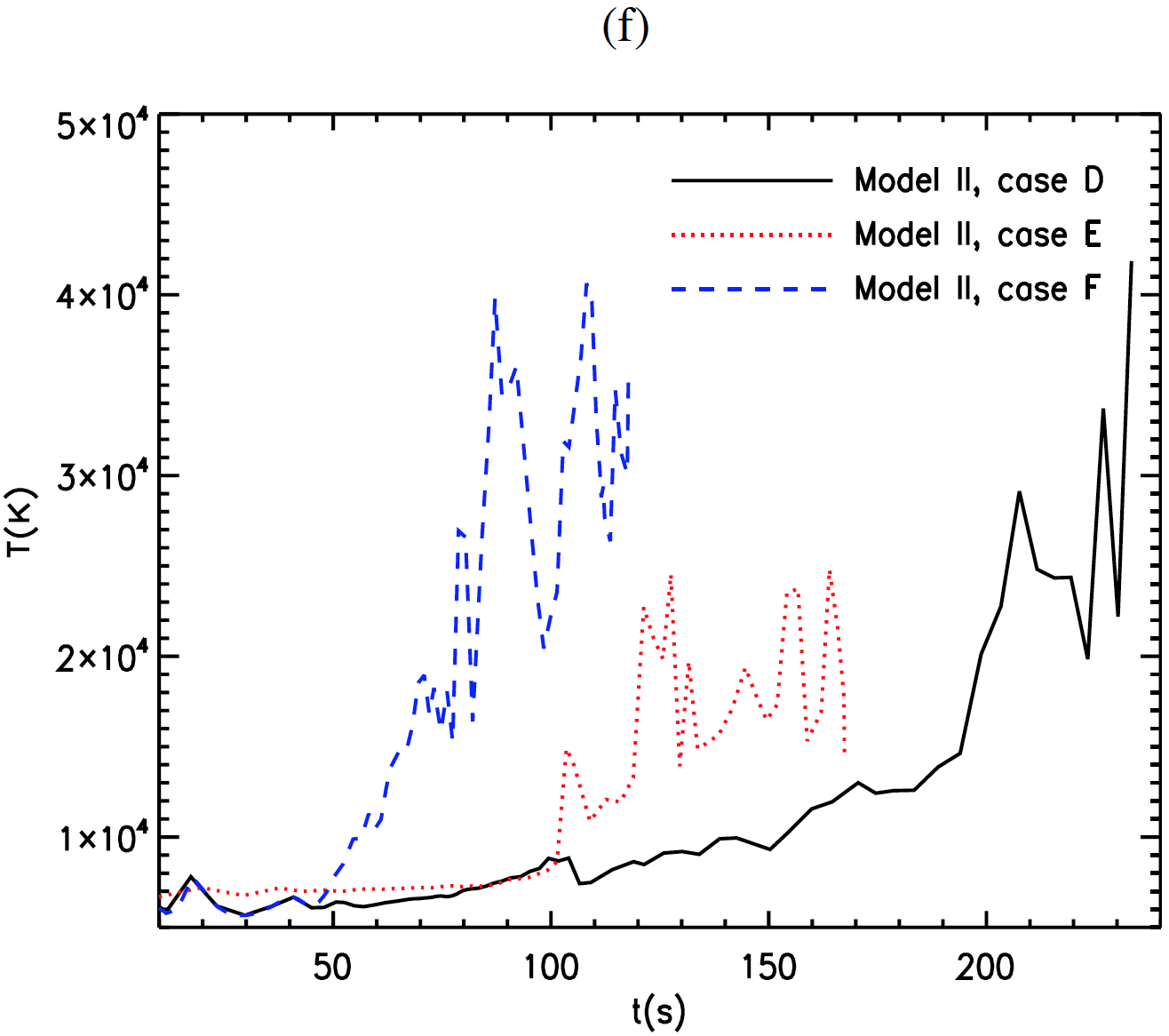}}
 \caption{Time evolution of the normalized reconnection rate $\gamma$, the current sheet width $\delta_{x}$ and the plasma temperature $T$ at the main X-point in different models and cases. Model I and model II represent the situations with strong guide field and zero guide field respectively. In each panel, the black line represents the case without radiative cooling and ambipolar diffusion, the blue line represents the case with ambipolar diffusion but without radiative cooling, the red line represents the case with radiative cooling but without ambipolar diffusion. Image and caption reproduced with permission from Ni \textit{et al}. (2015), copyright by AAS.}
 \label{fig8}
\end{figure}

\begin{figure}[!ht] 
 \centering\includegraphics[width=4in]{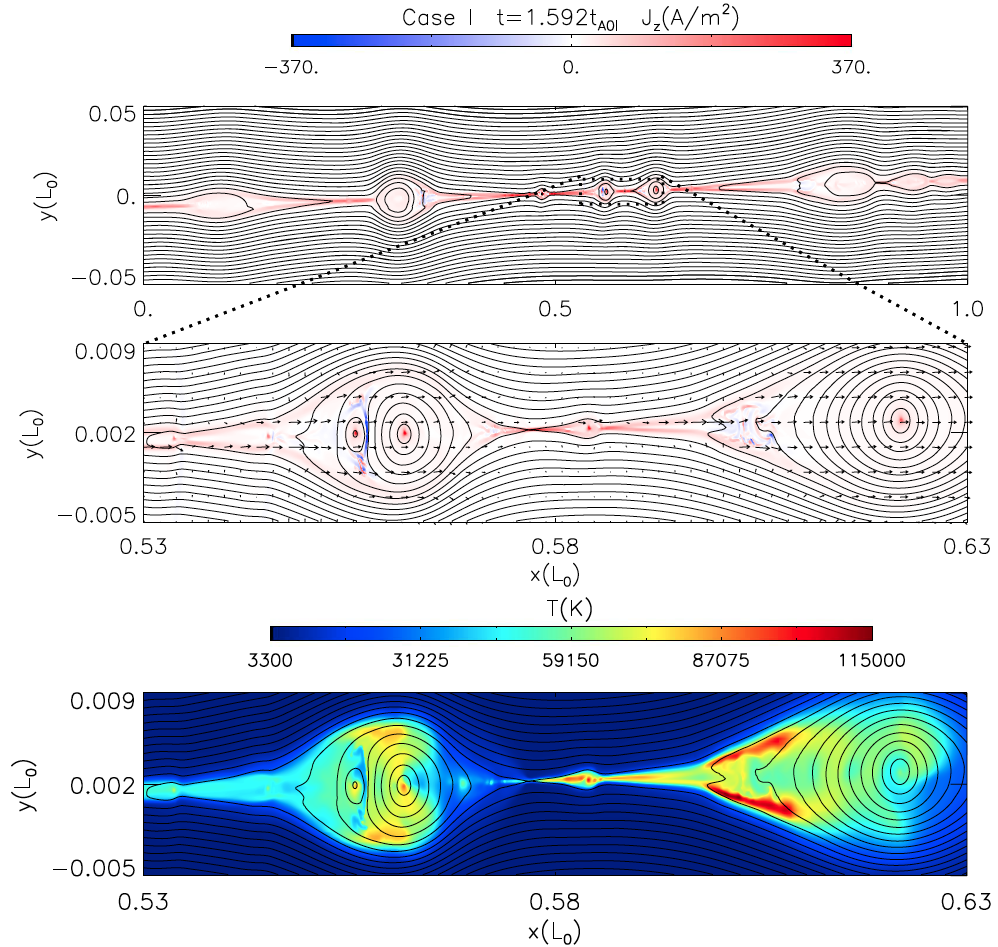}
 \caption{ The distributions of the current density $J_z$ and temperature $T$ after zoomed in to a small region with plasmoids. Image and caption reproduced with permission from Ni \textit{et al}. (2016), copyright by AAS  }
 \label{fig9}
\end{figure}

Figure~\ref{fig5}(a) indicates that ambipolar diffusion is the dominating diffusion mechanism above the TMR\@. Ambipolar diffusion represents the decoupling between the neutral particles and the ionized components. Though it does not change the magnetic topology during the reconnection process, it dissipates magnetic energy. The high resolution numerical studies about magnetic reconnection below the middle solar chromosphere have been performed\cite{ni:2015, ni:2016}. The simulations are based on the single-fluid MHD model, but the ambipolar diffusion and simple radiative cooling terms are included. The temperature dependent magnetic diffusion in the simulations is close to the Spitzer type as shown in equation 2.1. Figures~\ref{fig8} and \ref{fig9} show the results from these simulations\cite{ni:2015, ni:2016}. The three models with different distributions of initial magnetic field and gas pressure across the current sheet have been studied numerically\cite{ni:2015}. Model I and model II represent the situations with strong guide field and zero guide field respectively\cite{ni:2015}. The plasmoid instability cascading to smaller and smaller scales has been discovered in all the high resolution simulations, different orders of plasmoids appear in different length scales. The fourth-order is found as the highest order and the thinnest current width reaches about $30$~m\cite{ni:2015}. In Figure~\ref{fig8}(e), one can find that including the ambipolar diffusion makes the current sheet width to thin much faster before the plasmoids appear in model II with zero guide field\cite{ni:2015}, which is consistent with the previous analytical and one dimensional simulation results\cite{brandenburg:1994}. Figure~\ref{fig8}(d) and (f) show that the plasmoid instability and high temperature plasmas also appear earlier in the case by including ambipolar diffusion in model II with zero guide field. After plasmoid instability appears, the current sheet width at the main X-point, the reconnection rate and the temperature at the X-point eventually reaches the similar value in all the cases, no matter ambipolar diffusion is included or not\cite{ni:2015}. In model I with strong guide field, including ambipolar diffusion does not make any significant changes as shown in Figure~\ref{fig8}(a), (b) and (c), the current sheet width, the reconnection rate and the temperature at the main X-point vary with time similarly in all the cases in model I\cite{ni:2015}. These simulation results demonstrate that plasmoid instability is still one of the main mechanism to lead fast reconnection below the middle chromosphere. Plasmoid instability can also well explain the complex IRIS Si IV line profiles with bright cores and broad wings observed in the reconnection events in the low solar atmosphere\cite{innes:2015, rouppe:2017}. Ambipolar diffusion might play more important roles to result in fast reconnection and plasma heating in the upper chromosphere, which needs to be studied in further simulations. 

Figure~9 shows the distributions of the current density and temperature zoomed in to a small region including the plasmoids inside the current sheet, and this case represents a reconnection process with strong magnetic field ($B_0=500$~G, $\beta_0=0.058$) around the solar TMR\cite{ni:2016}. The simulation results show that the initial high density cold plasmas ($T_0=4200$~K) can be heated above $10^5$~K by small scale shocks inside the plasmoids in the reconnection process\cite{ni:2016}. The plasmoids in the reconnection region are multi-thermal. These simulation results indicate that UV bursts can be generated in a magnetic reconnection process with strong magnetic fields around the solar TMR\@, and the plasmas with different temperatures and densities can explain the different spectral lines from observations and they can appear at the same height. The multi-thermal plasmoids have been reported in several previous observations\cite{zhang:2016, zhang:2019}. The blob like fine structures in UV bursts have been uncovered by CHROMIS in Ca II K as discussed in the above subsection\cite{vissers:2019}.

The effect of recombination on magnetic reconnection in the chromosphere has been studied by adding an extra recombination term in the continuity equation based on the single-fluid MHD model\cite{sakai:1996}. The numerical results show that the recombination effect enhances the reconnection rate\cite{sakai:1996}. However, the ionization process and radiative cooling are not considered, which results in the unrealistic temperature increase in those simulations\cite{sakai:1996}. The momentum transfer and the frictions between the ionized plasmas and the neutrals have been included in the incompressible two-fluid (ion-neutral) MHD code to study magnetic reconnection in the low solar atmosphere\cite{ji:2001a, ji:2001b}. The authors did not show how the neutrals affect the reconnection process and only the sharp structures of magnetic fields and ions' velocity in the 3D simulations are presented\cite{ji:2001b}. Taking into account the transport processes of ion-neutral hydrogen collision, heat conduction, recombination and ionization, magnetic reconnection in the chromosphere has been further studied based on the compressible two-fluid MHD model\cite{smith:2008, sakai:2009}. The authors find that the dynamic evolutions of ions and neutrals in the two-fluid simulations are totally different from the results from the single-fluid simulations\cite{smith:2008}.  Their simulation results show that the reconnection rate for a fixed resistivity decreases as the plasmas become less ionized\cite{smith:2008}, suggesting that jets associated with fast reconnection must occur in the upper chromosphere\cite{smith:2008}, where the ionization degree is higher. They also find that the reconnection rate is not increased by including the ionization and recombination effects\cite{smith:2008}. However, the collision frequency, ionization and recombination rates did not depend self-consistently on the plasma temperature and density in their simulations, therefore do not vary with time and space during the reconnection process. In addition, the fixed resistivity in their simulations is about four orders of magnitude larger than a typical value in chromosphere, the plasmoid instability then does not appear.     
 
\begin{figure}[!ht]
 \centering\includegraphics[width=4in]{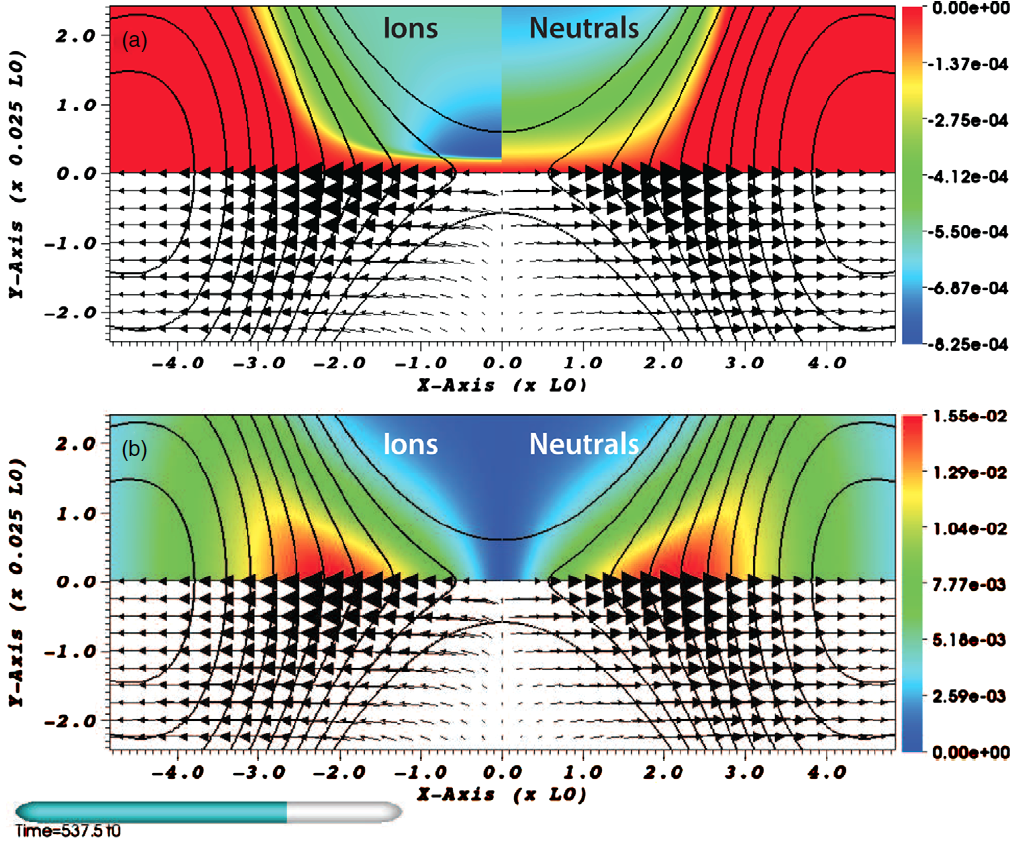}
 \caption{Plots of ion and neutral flow showing inflow decoupling and outflow coupling during reconnection. Panel (a) shows color contours of the dimensionless vertical velocity $(v_{i,y})/v_0$ for the ions (top left quadrant) and for the neutrals ($v_{n,y}/v_0$, top right quadrant). The solid  contour lines show 10 values of the magnetic flux $A_z$, regularly distributed in the interval $[-0.04, -0.01]B_0L_0$. The arrows on the bottom left quadrant represent the plasma flow, and those on the bottom right quadrant the neutral flow. Panel (b) shows the same as panel (a) but for the magnitude of horizontal velocity for ions ($v_{i,x}/v_0$) and neutrals ($v_{n,x}/v_0$). Image and caption reproduced with permission from Leake \textit{et al}. (2012), copyright by AAS.}
 \label{fig10}
\end{figure}

\begin{figure}[!ht]
 \centering\includegraphics[width=4in]{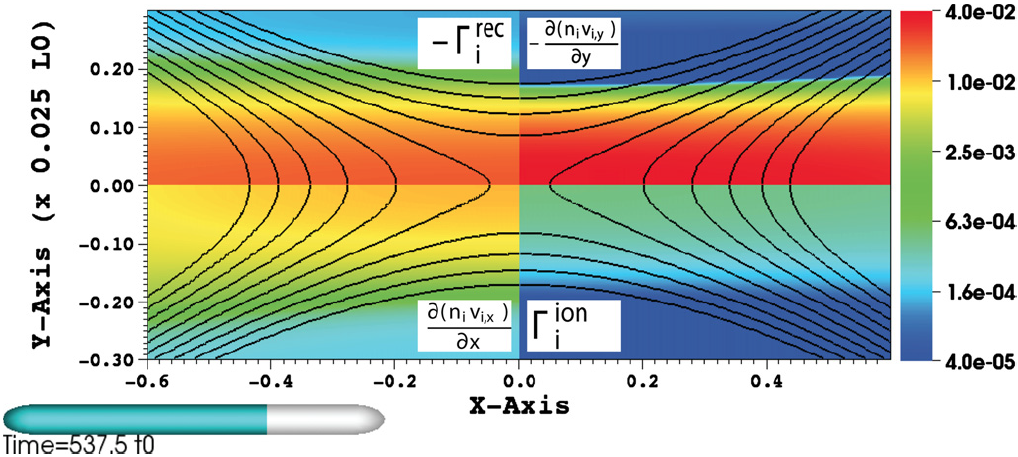}
 \caption{Steady state reconnection region showing contributing sources and sinks of ions in the current sheet (in units of $L_0^{-3}t_0^{-1}$). Top left quadrant shows rate of loss of ions due to recombination. The bottom left quadrant shows rate of loss of ions due to outflow $\partial{(n_i v_{i,x})}/{\partial x}$. Top right quadrant shows rate of gain of ions due to inflow $-(\partial{(n_i v_{i,y})}/{\partial y})$. The bottom right quadrant shows rate of gain of ions due to ionization. The solid  contour lines show 10 values of the magnetic flux $A_z$, evenly distributed in the interval $[-0.0378, -0.037]B_0L_0$. This shows that loss of ions due to recombination and outflow is comparable, and combine to balance the inflow of ions, with ionization playing an insignificant role. Image and caption reproduced with permission from Leake \textit{et al}. (2012), copyright by AAS  }
 \label{fig11}
\end{figure}

\begin{figure}[!ht]
 \centering\includegraphics[width=4in]{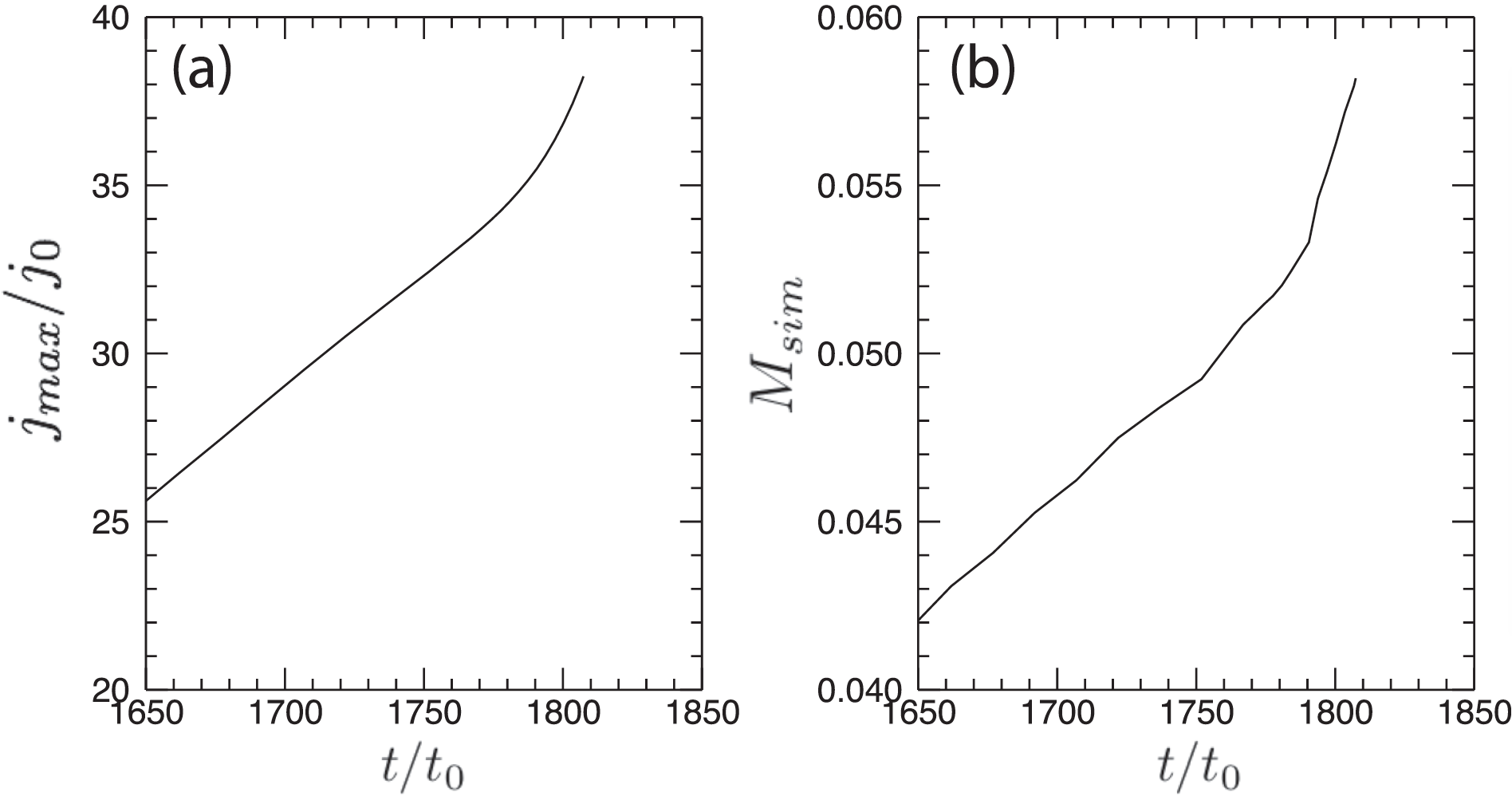}
 \caption{Temporal evolution of the reconnection rate during the latter stages when the plasmoid instability sets in. The plasmoids are not observed until $t=1790t_0$. Left panel: $j_{max}/j_0$, the maximum value of the out of plane current density. Right panel: $M_{sim}=\eta^{\ast} j_{max}/(V_{A}^{\ast}B_{up})$, the normalized reconnection rate. $j_{max}$ is located at the main reconnection X-point in this work. Image and caption reproduced with permission from Leake \textit{et al}. (2013), copyright by AIP.}
 \label{fig12}
\end{figure}

 The first self-consistent 2D simulations of chromospheric magnetic reconnection based on the reactive multi-fluid plasma-neutral module\cite{meier:2012} are performed\cite{leake:2012, leake:2013, murphy:2015} within the HiFi modeling framework\cite{lukin:2008}, ambipolar diffusion is then self-consistently included. The simulations also include ion-neutral scattering collisions, ionization, recombination, optically thin radiative loss, collisional heating, and thermal conduction.  The initial conditions represent that magnetic reconnection happens around the middle chromosphere, the reconnection magnetic fields are only about $10-50$~G and the initial plasma $\beta$ is larger than one. The numerical results show that the neutral and ion fluids become decoupled upstream from the reconnection site\cite{leake:2012}, creating an excess of ions in the reconnection region. Figure~\ref{fig10} clearly shows that the decoupling of ions and neutrals appear in the upstream region, but they are coupled well in the out flow region. Figure~\ref{fig11} shows that the ion recombination inside the current sheet is much stronger that the neutral ionization. Ion recombination in the reconnection region, combined with Alfvenic outflows, quickly removes ions from the reconnection site\cite{leake:2012}, leading to a fast reconnection independent of Lundquist number\cite{leake:2012}.  Their further studies with more realistic chromospheric transport coefficients and magnetic diffusion proved this point again\cite{leake:2013}. Figure~\ref{fig12} shows that the strong recombination and outflows already make the reconnection rate to reach above $\sim0.05$ before the plasmoid instability appears, the contribution of plasmoid instability on the fast reconnection rate is small\cite{leake:2013}.  The temperature increases in these simulations are only about several hundred to several thousand Kelvins.
 
\begin{figure}[!ht]
  \centerline{\includegraphics[width=1.8in]{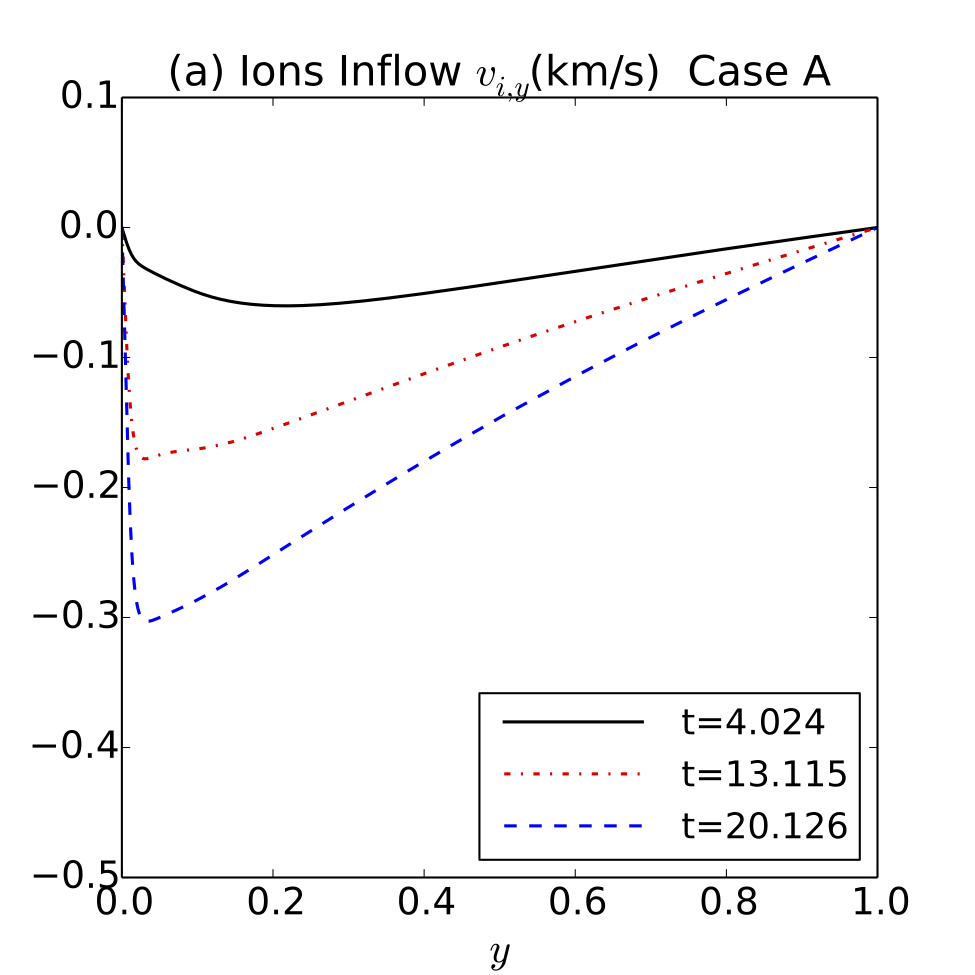}
                       \includegraphics[width=1.8in]{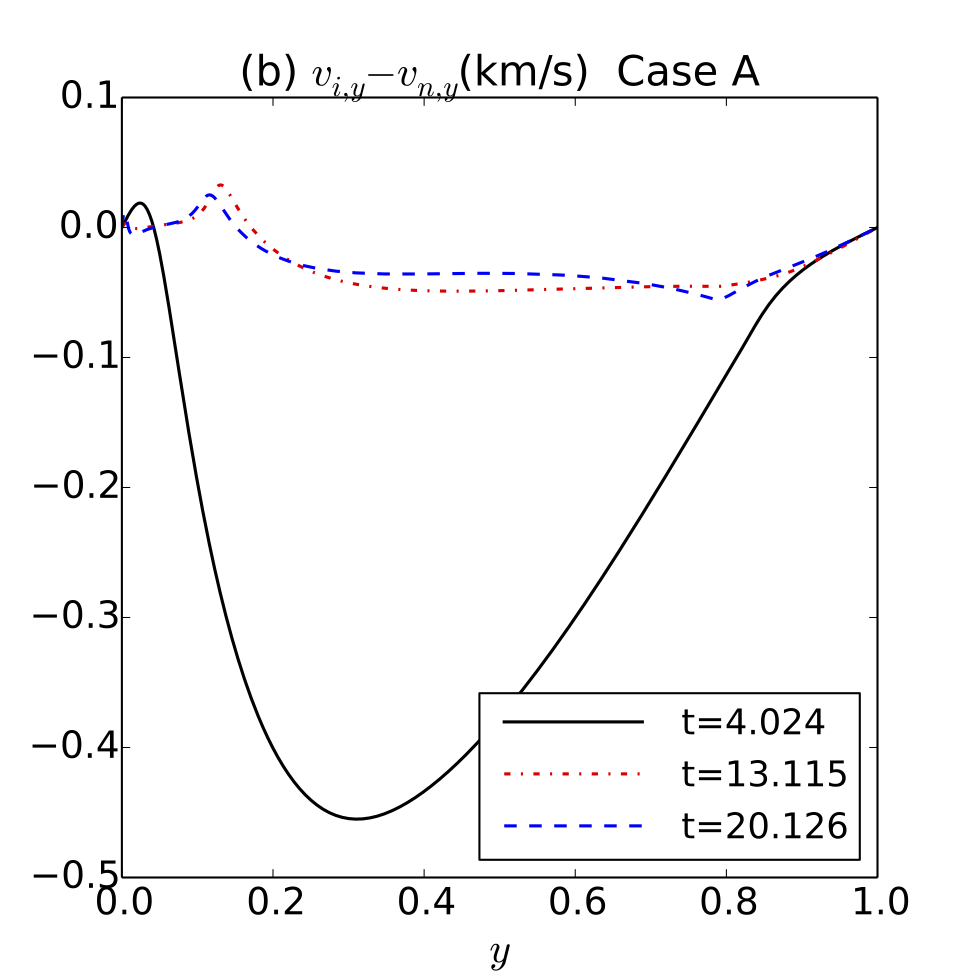}
                       \includegraphics[width=1.8in]{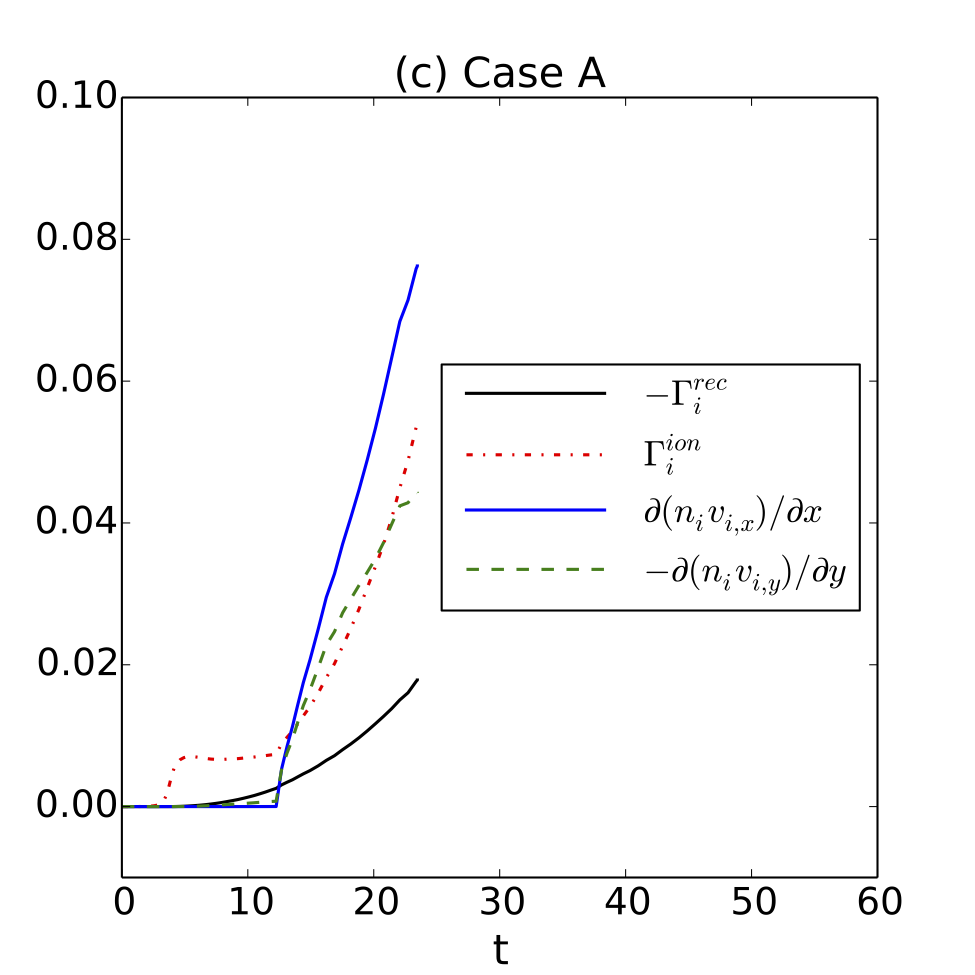}}
 \caption{Panels (a) and (b) show the ion inflow speed $v_{i,y}$, and the difference in speed between the ion and the neutral inflows $v_{i,y}-v_{n,y}$ across the current sheet at $x=0$ at $t=4.024$, $t=13.115$, and $t=20.126$ in Case~A. Panel (c) shows the time-dependent contributions to $\partial{n_i}/\partial{t}$ in Case A. The four contributions are the average values inside the current sheet, the loss due to recombination $-\Gamma_{i}^{rec}$, the loss due to the outflow $\partial{(n_i v_{i,x})}/\partial x$, the gain due to the inflow $-\partial{(n_i v_{i,y})}/\partial y$, and the gain due to ionization $\Gamma_i^{ion}$. In Case~A, the initial strength of the magnetic field is $B_0=500~G$ and the initial plasma $\beta$ including ions and neutrals is $\beta_0=0.058$.The time $t$ is in dimensionless. This shows that the effect of ionization is always larger than the recombination during the reconnection process in such a low $\beta$ case. Image and caption reproduced with permission from Ni \textit{et al}. (2018a), copyright by AAS.}
 \label{fig13}
\end{figure}

Since UV bursts are usually formed in active regions, the reconnection magnetic fields can be as high as several hundred or thousand Gauss. In order to study the generation mechanisms of UV bursts, magnetic reconnection around the solar TMR region has been studied \cite{ni:2018a, ni:2018b, ni:2018c} by using the same reactive multi-fluid plasma-neutral module within the HiFi modeling framework, the strength of reconnection magnetic fields is much stronger than those in the previous studies\cite{leake:2012, leake:2013, murphy:2015}. In such low $\beta$ and strong magnetic field cases, the numerical results show that the ionized and neutral fluid flows are well coupled throughout the whole reconnection region\cite{ni:2018a, ni:2018b, ni:2018c}, which is clearly demonstrated by comparing Figure~\ref{fig13}(a) and \ref{fig13}(b). Ionization rate is always larger than recombination rate in the reconnection region\cite{ni:2018a} as shown in Figure~13(c). Recombination effect plays a small effect on accelerating the reconnection rate\cite{ni:2018a, ni:2018b, ni:2018c} in this case. Since the strong ionization consumes a large amount of energy, the temperature increase in a reconnection process by including the non-equilibrium ionization effect\cite{ni:2018a} is much smaller than that in the previous one-fluid simulations. However, the maximum temperature exceeds $2\times10^4$~K when the initial plasma  $\beta$ is lower than $0.058$ and the strength of reconnection magnetic field is larger than $B_0=500$~G\cite{ni:2018a, ni:2018b}. Figure~14 shows the distributions of the ionization degree $f_i$ and the ion temperature $T_i$ inside the current sheet region at the end of the simulations in four cases with different strengths of magnetic fields. Though the plasma is not heated above $8\times10^4$~K as shown in the previous one-fluid simulation\cite{ni:2016}, the Si IV emission line is still possibly to be formed when the temperature for such a high density plasma reaches about $2\times10^4$~K\cite{rutten:2016}. The reconnection processes at different heights through the photosphere to the low chromosphere have also been studied\cite{ni:2018c}. The simulations show that it is difficult to heat the much denser photospheric plasmas to above $2\times10^4$ K during the magnetic reconnection process\cite{ni:2018c}. However, the plasmas in the low solar chromosphere can be heated above $3\times10^4$ K with reconnection magnetic fields of 500 G or stronger\cite{ni:2018c}. These multi-fluid simulations based on the HiFi modeling framework are eventually always terminated because of the grid errors caused generally by insufficient resolution\cite{ni:2018a, ni:2018b, ni:2018c}, the temperatures can possibly be increased to higher values if the simulations can last for longer time.
 
\begin{figure}[!ht]
  \centerline{\includegraphics[width=2.0in]{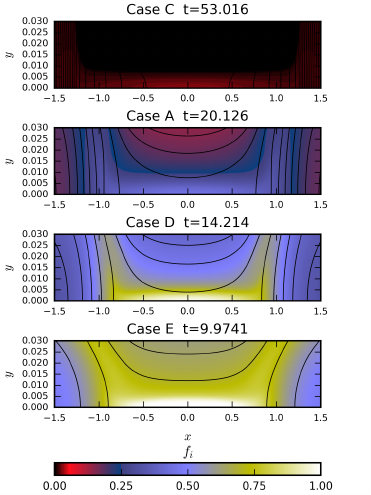}
                       \includegraphics[width=2.0in]{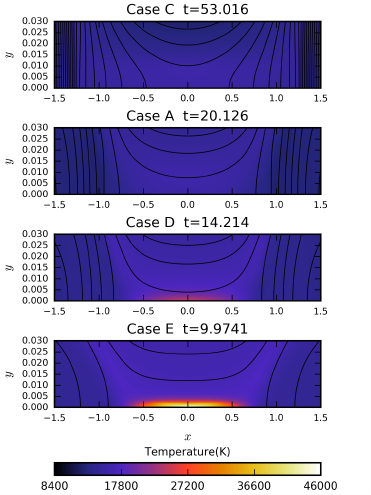}}
 \caption{ Left panel and the right panel respectively show the distributions of the ionization degree $f_i$ (represents the same meaning as the ionization fraction $\chi$ in section 2) and the neutral temperature $T_i$ inside the current sheet region in half of the domain, at the end of the simulations in four different cases. The initial strengths of magnetic fields and plasma $\beta$ are $B_0=100$\,G and $\beta_0=1.46$ in Case C, $B_0=500$\,G and $\beta_0=0.058$ in Case A, $B_0=1000$\,G and $\beta_0=0.0145$ in Case D, and $B_0=1500$\,G and $\beta_0=0.0064$ in Case E\@. The characteristic length scale is $L_{\star}=100$\,m in all the four cases. Image and caption reproduced with permission from Ni \textit{et al}. (2018a), copyright by AAS.}
 \label{fig14}
\end{figure} 

\begin{figure}[!ht]
  \centerline{\includegraphics[width=2.6in]{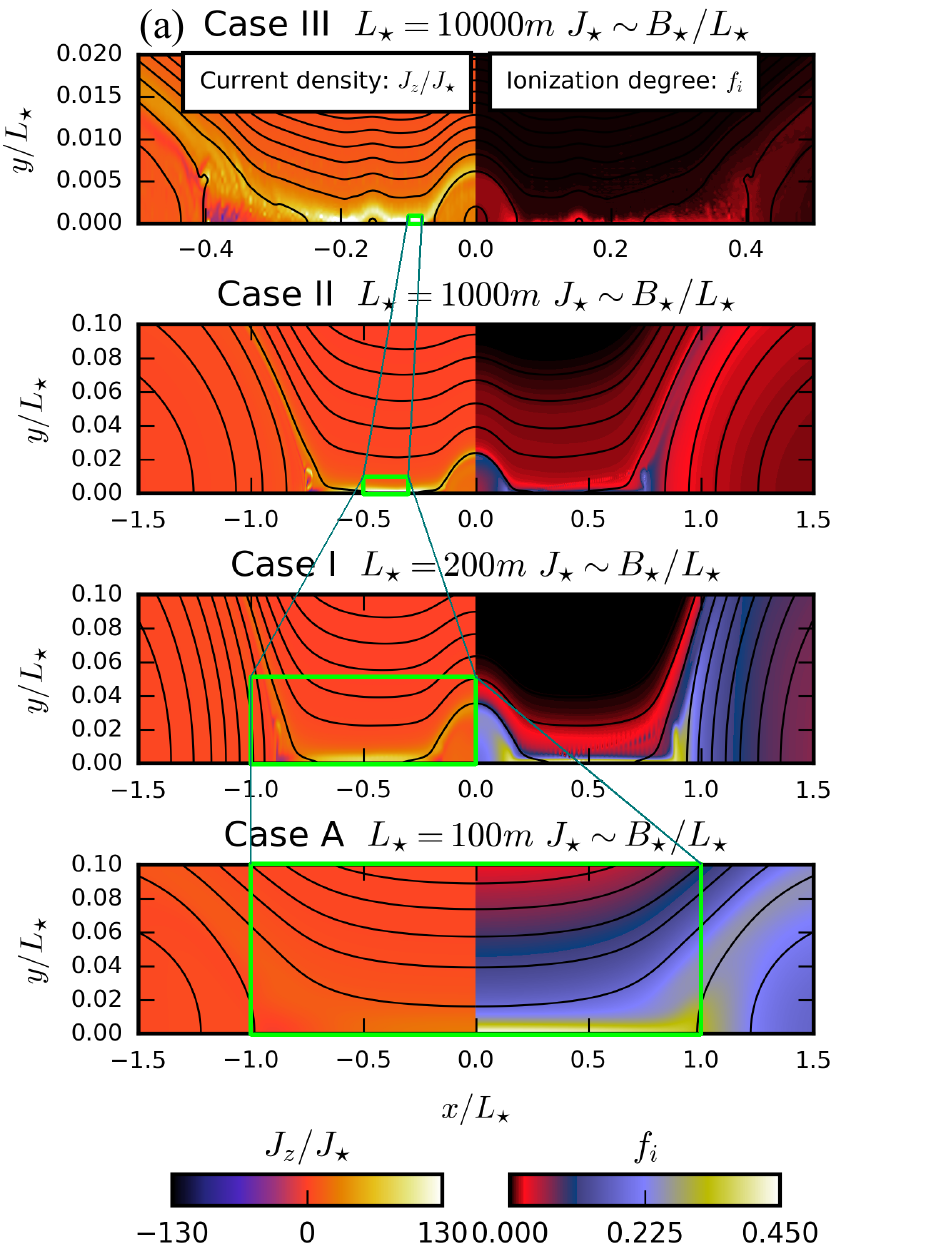}
   \includegraphics[width=2.6in]{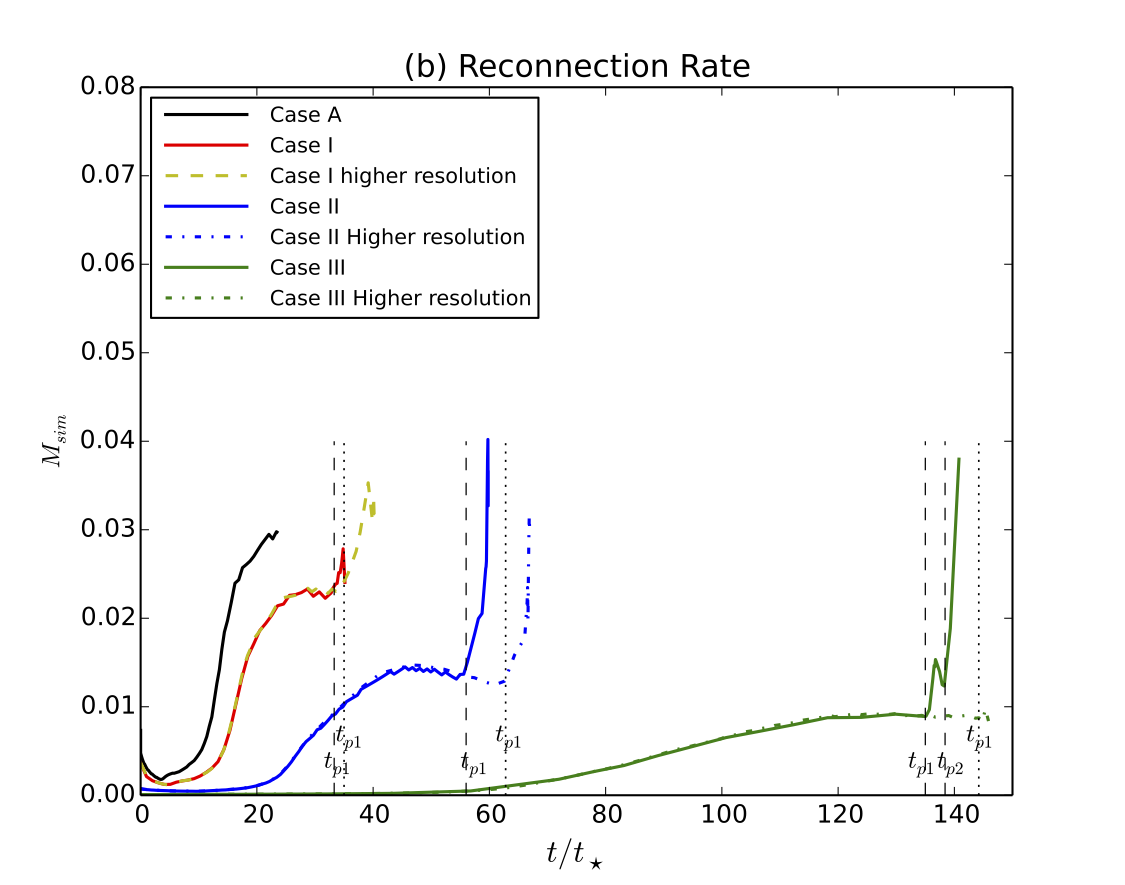}}
 \caption{ (a) shows magnetic reconnection during the later stage at different length scales. The left side of panel (a) shows the distributions of the dimensionless current density $J_z/J_{\star}$ and the right side of panel (a) shows the ionization fraction $f_i$. (b) shows the calculated reconnection rate $M_{sim}$ vs.\ time in Case~A, Case I, Case II and Case III\@. Plasmoid instability appear in all the cases except in Case A\@. $t_{p1}$ represents the time point when the first magnetic island appeared inside the current sheet in each run, $t_{p2}$ represents the time point when the second magnetic island appeared inside the current sheet in the lower resolution run of Case III\@. The values of the characteristic length scale $L_{\star}$ are $100$\,m in Case A, $200$\,m in Case I, $1000$\,m in Case II and $10000$\,m in Case III\@. The values of the characteristic time $t_{\star}$ are $0.0029$\,s in Case A, $0.0058$\,s in Case I, $0.029$\,s in Case II and $0.29$\,s in Case III\@. The other characteristic values are the same in all the four cases, except for the length scale and the ones relating with length scale. This figure shows that the plasmoid instability eventually terminates in a scale with length around 100 m and width around 2 m. Image and caption reproduced with permission from Ni \textit{et al}. (2018b), copyright by AAS.}
 \label{fig15}
\end{figure}

The low $\beta$ multi-fluid simulations within the HiFi modeling framework have also been performed within different length scales. The larger length scale means the higher Lundquist number in these simulations. Figure~\ref{fig15}(a) shows that the current sheet is more turbulent and more plasmoids appear in the case with a larger length scale and higher Lundquist number. Figure~\ref{fig15}(b) shows the time dependent reconnection rate in all the four cases simulated with different resolutions. The reconnection rate in Case~III is the lowest in all cases before the plasmoid instability appears, with the reconnection rate approximately scaling with the Lundquist number as $M_{sim}\sim S_{sim}^{-1/2}$ at $t_{p1}$, where $S_{sim}$ is the Lundquist number calculated from the simulations. After the second magnetic inland appears in the lower resolution run of Case~III, the reconnection rate sharply reaches above $0.035$, resulting in a five-fold increase in the reconnection rate during a very short period. As we know from observations, the reconnection scale in a UV burst is usually about $1$\,Mm, which is much larger than that in Case~III\@. Therefore, we can conclude that plasmoid instability is the main mechanism to lead fast reconnection in the case with strong magnetic fields below the middle chromosphere.

Figure~\ref{fig5}(a) shows that the Hall effect could possibly be important above the solar TMR\@. The onset of fast Hall reconnection in partially ionized plasmas has been analytically studied based on the incompressible three-component plasma module by ignoring the ionization and recombination processes\cite{malyshkin:2011}.  The analytical results show that the width of the Sweet-Parker layer far exceeds the ion inertial length in the solar chromosphere\cite{malyshkin:2011}, which means that the fast Hall-mediated reconnection is unfavorable\cite{malyshkin:2011}. Using the reactive multi-fluid plasma-neutral module within the HiFi modeling framework, the Hall effect has been included in the induction equation to study magnetic reconnection around the middle chromosphere\cite{murphy:2015}. The numerical results show that the signatures of the Hall effect-generated magnetic fields are clearly evident at $d_i^{\prime}=d_i\sqrt{\rho/\rho_i}$ spatial scale\cite{murphy:2015}. However, the reconnection rate acceleration by Hall effect is not observed in their simulations\cite{murphy:2015}. The Hall effect on magnetic reconnection with strong magnetic fields around the solar TMR has also been tested in weakly ionized plasmas\cite{ni:2018a}. The numerical results show that Hall effect only slightly increases the reconnection rate at beginning\cite{ni:2018a}, but in the case with guide field it does not result in significant asymmetries or change the characteristics of the reconnection current sheet down to meter scale\cite{ni:2018a}. As shown in Figure~\ref{fig15}(a), a series of simulations at different scales demonstrate a cascading current sheet formation process that terminates for current sheets with width of $2$~m and length of $\sim100$~m\cite{ni:2018c}, corresponding to the critical current sheet aspect ratio of $\sim50$.  Since the initial weakly ionized plasmas become highly ionized during the later stage of the reconnection process with strong magnetic fields, the density of the ionized plasmas have been increased by $10^3 \sim 10^4$ times, which results in the strong decrease of the ion inertial length and the ion-neutral collision mean free path. The width of the current sheet of $2$~m is still about $1\sim 2$ orders of magnitude larger than the ion-inertial length and ion-neutral collision mean free path inside the current sheet. Therefore, the plasmoid cascading terminates in MHD scale\cite{ni:2018c}. These previous results demonstrate that Hall effect is not important in a magnetic reconnection process below the middle chromosphere\cite{ni:2018a, ni:2018c, murphy:2015}. However, the Hall effect on magnetic reconnection in the upper chromosphere can be important. The recent kinetic simulations of magnetic reconnection in partially ionized upper chromosphere show that the transition to fast reconnection occurs when the current sheet width thins below the ion inertial length\cite{jara:2019}.  
 
 Radiative cooling is very important in the low solar atmosphere because of the high density plasmas. The previous analytical results demonstrate that the strong radiative cooling also results in the quick thinning of the current sheet and a much faster reconnection rate when the guide field is zero\cite{uzdensky:2011}. As shown in Figure~\ref{fig8}, including the radiative cooling indeed makes the current sheet thins quicker and the fast reconnection appears earlier in the case with zero guide field in the single-fluid simulations\cite{ni:2015}. However, the heating term included to balance the initial cooling term makes the radiative cooling effect not that obvious as the ambipolar diffusion effect\cite{ni:2015}. Two different radiative cooling models have been applied in the reactive multi-fluid plasma-neutral module within the HiFi modeling framework to study reconnection around the solar TMR\cite{ni:2018b}. The first simple radiative cooling model represents the radiative losses that are due primarily to radiative recombination, with a very crude approximation for radiation due to the presence of excited states of neutral hydrogen\cite{ni:2018b}. The more realistic radiative cooling model computed using the OPACITY project and the CHIANTI databases is also applied\cite{goodman:2012}. Most of the generated thermal energy is radiated in the reconnection process, no matter which radiative cooling model is used\cite{ni:2018a, ni:2018b}. Though the main conclusions are not changed by using the two different radiative cooling models\cite{ni:2018a, ni:2018b}, the time and space evolutions of temperatures and ionization degree depend on the radiative models\cite{ni:2018a, ni:2018b}. These two radiative cooling models have also been applied in another code with reactive multi-fluid plasma-neutral module to study chromosphere magnetic reconnection with weak magnetic field\cite{alvarez:2017}. The numerical results show that radiative cooling changes the plasma pressure and the concentration of ions inside the currents sheet, the strong cooling produces faster reconnection than has been found in models without radiation\cite{alvarez:2017}. The photosphere is optical thick and the plasmas usually obey the local thermodynamic equilibrium (LTE). The solar chromosphere radiative losses are much more complicated, and the plasmas there do not obey LTE or statistical equilibrium. Therefore, the expensive full radiative transfer computations are necessary when a simulation covers the solar chromosphere. The chromosphere radiative transfer process in the reconnection events in the low solar atmosphere has been well computed in the previous simulations\cite{hansteen:2017, hansteen:2019} by using the Bifrost code\cite{gudiksen:2011}.

\section{Reconnection in partially ionized astrophysical plasmas\label{astrophysics}}

Magnetic reconnection occurs in a variety of partially ionized astrophysical plasmas beyond the solar chromosphere.  Studying magnetic reconnection in partially ionized astrophysical plasmas outside of the heliosphere is difficult because of the inability to spatially resolve the reconnection region.  In this section, we discuss the nature of magnetic reconnection in stellar chromospheres, the predominantly neutral phases of interstellar plasmas, and protostellar/protoplanetary disks.

\subsection{Stellar chromospheres\label{stellar}}

Chromospheres are ubiquitous around main sequence stars ranging from middle F type to late M type \cite{linsky:2017}.  Chromospheres have been observed around main sequence stars with an effective temperature of up to 8250 K\@.  At the cool end of the main sequence, chromospheres have been observed around some early-type L stars.  Chromospheres have been observed around some pre-main sequence stars (including weak-line T Tauri stars) and around some post-main sequence stars.  Many of these stars are magnetically active and demonstrate chromospheric activity \cite{hall:2008}.  While evidence for magnetic reconnection in transition region plasma has been reported in stars besides the sun \cite{linsky:1994}, we know of no direct observations of reconnection in the partially ionized chromospheres of other stars.

The properties of chromospheres vary significantly from star to star \cite{linsky:2017}.  The overall structure of a chromosphere depends on a star's spectral type, elemental abundances, convection zone properties, radiation field, and magnetic field properties.  The magnetic field of a star depends in turn on stellar rotation and age.  Physical conditions throughout stellar chromospheres vary significantly between stars.  For example, the temperature minimum is typically between roughly 2000 K and 5000 K \cite{linsky:2017}.  Chromospheric structure and variation affects plasma parameters such as the ionization fraction, neutral-ion mean free path, and ion inertial length.  The scaling properties of reconnection in different stellar chromospheres remains to be studied in detail.

\subsection{Neutral phases of the interstellar medium\label{ISM}}

The interstellar medium (ISM) includes multiple phases \cite{ferriere:2001, cox:2005, draine:2011}.  The hot ionized medium (HIM) is composed of plasmas that have been heated to 
{$\sim$}{$10^5$} to {$\sim$}{$10^7$} K\@ by supernova blast waves. The warm ionized medium (WIM) is predominantly composed of hydrogen at $T \sim 8000$~K that has been photoionized by radiation from hot stars.  The warm neutral medium (WNM) is predominantly composed of atomic hydrogen at temperatures of $\sim$6000~K\@.  The cold neutral medium (CNM) is largely atomic hydrogen at temperatures of $\sim$100~K\@.  Molecular clouds (which are discussed in the next subsection)
are predominantly composed of dense H$_2$ at temperatures of $\sim$20~K and are host to star formation.  Roughly 1\% of the mass of the ISM is dust, which often becomes charged.  Highly energetic cosmic rays are prevalent throughout the ISM\@.  The thermal energy density, turbulent kinetic energy density, magnetic energy density, and cosmic ray energy density are all within an order of magnitude of each other.  The ionization fractions of the WNM, CNM, and molecular clouds are substantially greater than would be expected from collisional ionization for a given temperature.  Ionization in the neutral phases of the ISM is predominantly due to photoionization from starlight and collisions between cosmic rays and neutral atoms or molecules.  In star-forming clouds, photoionization of lithium by thermal photons trapped within the cloud may be the dominant mechanism \cite{nakauchi:2019}.  The predominantly neutral phases of the ISM account for perhaps a third of the volume and the majority of the mass.  Drivers of turbulence in the ISM include supernovae, protostellar outflows, hot star winds, and instabilities.

The magnetic fields of spiral galaxies tend to be ordered on large scales.  For example, there is observational evidence for a magnetic field reversal in the Sagittarius-Carina arm of the Milky Way \cite{han:1994}.  The ordered component of the galactic magnetic field suggests that magnetic reconnection occurs efficiently in the ISM\@.  Moreover, reconnection is thought to be necessary in order for the galactic dynamo to occur.  The multiscale nature of magnetic reconnection is especially apparent for interstellar plasmas, as the vast length scales correspond to Lundquist numbers of up to $\sim${$10^{20}$}.  A satisfactory description of reconnection in the predominantly neutral phases of the ISM must include not just a mechanism for fast reconnection, but also an explanation for how current sheets are able to form in the first place.  

One of the leading theories of the formation of current sheets in the interstellar medium is thinning due to ambipolar diffusion \cite{brandenburg:1994, brandenburg:1995, zweibel:1997, heitsch:2003a}, as discussed in Section \ref{theories}.  In fluid models, ambipolar diffusion can lead to a singularity in the current density as a result of this thinning process.  Pressure and resistivity can remove the singularity, but non-fluid effects such as a finite ion gyroradius will provide more stringent constraints on current sheet thinning for parameters characteristic of the ISM \cite{brandenburg:1995}.  Turbulence enables faster ion-neutral drift \cite{zweibel:2002}.

Turbulence is strongly believed to play key roles in reconnection in all phases of the ISM\@.  Lazarian \& Vishniac \cite{lazarian:1999} presented a model for magnetic reconnection in weakly stochastic magnetic fields.  They argued that a reconnection region consists of multiple smaller scale reconnection sites, and that field line wandering determines the reconnection rate rather than any particular dissipation mechanism.  They therefore predicted that reconnection will be fast in a turbulent medium even when collisional resistivity is minuscule.  A formal mathematical treatment of the problem of Richardson diffusion recovers the scalings from this model \cite{eyink:2015}.  Numerical simulations by later members of this group have been consistent with many of the predictions \cite{kowal:2009, kowal:2012}. This model does not include kinetic effects \cite{karimabadi:2013}, which are observed to play key roles during reconnection in laboratory and heliospheric plasmas. We refer the reader to the review by Lazarian et al.\ \cite{lazarian:2020} which discusses 3D turbulent reconnection models in substantially more detail.

The original model of Lazarian \& Vishniac \cite{lazarian:1999} only considered reconnection in a fully ionized, turbulent plasmas.  Lazarian, Vishniac, \& Cho \cite{lazarian:2004} later extended this model to describe stochastic reconnection in a partially ionized plasma.  In a partially ionized plasma, neutral particle viscosity and ion-neutral collisional coupling provide additional damping of fluctuations that does not occur in fully ionized plasmas.  These authors provided arguments that the turbulent cascade continues below the viscous cutoff scale.  They predicted a reduction in the reconnection rate due to the presence of neutrals, but only by an order of magnitude.  Later two-fluid simulations \cite{kowal:2012} found that the ions and neutrals followed different cascades below the ambipolar diffusion scale if the turbulence is globally super-Alfv\'enic.  The ion cascade continued below that scale because ions were decoupled from neutrals.  When the turbulence is globally sub-Alfv\'enic, turbulence is damped at the ambipolar diffusion scale.  These simulations suggest that reconnection diffusion \cite{lazarian:1999, eyink:2015, lazarian:2004, lazarian:2014} due to turbulence is plausible in portions of the ISM such as giant molecular clouds where turbulence is super-Alfv\'enic.  Because observational constraints on reconnection in the ISM are very limited, it is not known whether ambipolar diffusion, turbulence, or a combination of the two processes is necessary to describe reconnection in the CNM or WNM\@.

A longstanding problem in star formation theory is how mass can bypass the magnetic field in order to allow contraction and collapse of a molecular cloud core.  Potential solutions to this problem generally invoke some combination of ambipolar diffusion, turbulence, and reconnection.  Crutcher \cite{crutcher:2012} states that there is no definitive evidence in favor of ambipolar-diffusion-driven star formation, and suggests that reconnection plays a stronger role than ambipolar diffusion.  Lazarian \cite{lazarian:2014} argues that reconnection diffusion should be the primary mechanism and states that the reconnection diffusion and ambipolar-diffusion-driven star formation models provide different observational predictions.  When the ionization fraction is large, ambipolar diffusion models predict that the plasmas and magnetic fields will be tightly coupled.  In contrast, reconnection diffusion depends on the scales of turbulent eddies and the amplitude of turbulence.  

\subsection{Protostellar and protoplanetary disks\label{disks}}

Circumstellar disks form around protostars as a consequence of angular momentum conservation during star formation \cite{wurster:2018}.  The plasmas in these disks accrete into the protostar, forms planets, or gets ejected via winds and jets.  Models of protostellar and protoplanetary disks must explain not just how rotationally supported disks form, but also how angular momentum is transported outward so that matter can accrete into the protostar.  Magnetic reconnection is believed to play important roles in disk formation and evolution.  Resistivity, ambipolar diffusion, and the Hall effect are expected to play different roles in different parts of protostellar disks.

The gas and plasmas in protostellar and protoplanetary disks are typically dense and weakly ionized \cite{wurster:2018}.  The dominant ionization mechanism in these disks is collisions between neutrals and cosmic rays.  Cosmic rays will be attenuated when passing through a dense medium, so the inner regions of the disk near the midplane may be shielded from cosmic ray ionization \cite{padovani:2014}.  The regions of the disk that face the protostar may be photoionized by X-rays \cite{igea:1999}, and additional ionization may occur from stellar energetic particle events (which are analogous to solar energetic particle events).  Near the midplane, radioactive decay of unstable isotopes may be the dominant ionization mechanism.  The ionization fraction could be as low as $\sim${$10^{-14}$} \cite{wurster:2018}.  If the ionization fraction becomes too low, the material in the disk may act as an unmagnetized gas rather than a magnetized plasma so that the concept of reconnection no longer applies.

The formation of a rotationally supported disk is suppressed in idealized simulations with a strong enough initial magnetic field that is parallel to the rotation axis of the central object \cite{wurster:2018, allen:2003, Mellon:2008}.  This problem is known as the magnetic braking catastrophe.  Misalignment between the magnetic field and rotation axis \cite{chapman:2013, hull:2013, krumholz:2013, li:2013, gray:2018}, turbulence \cite{gray:2018, santoslima:2012, santoslima:2013, joos:2013, seifried:2013, simon:2015, lam:2019}, and non-ideal effects \cite{lam:2019, mellon:2009, krasnopolsky:2010, krasnopolsky:2011, dapp:2010, dapp:2012, bai:2014, tsukamoto:2015, tsukamoto:2017, wurster:2016, wurster:2019} are being investigated as interrelated solutions to the magnetic braking catastrophe.  These effects likely work in conjunction with each other.  The dissipation range of turbulence depends on non-ideal effects, and turbulence can change the alignment of the magnetic field.  The required resistivity to enable rotationally supported disk formation is a few orders of magnitude larger than the microscopic resistivity \cite{krasnopolsky:2010}.  Santos-Lima et al. \cite{santoslima:2012} argue that reconnection diffusion enabled by turbulence provides enough magnetic flux removal to enable rotationally supported disk formation.  Li et al. \cite{li:2014} present a scenario where a pseudodisk is warped out of the equatorial region.  Plasmas out of the equatorial region may undergo gravitational collapse, which can severely pinch the magnetic field and lead to reconnection.  

A longstanding problem in astrophysics is how material in accretion disks can accrete into a protostar on reasonably short time scales, as viscosity by itself cannot explain fast accretion.  The magnetorotational instability (MRI) occurs when a weak magnetic field is present in an accretion disk with Keplerian flow \cite{balbus:1991, balbus:1998}. This instability induces radial flows, excites turbulence, and transports angular momentum.  Magnetic reconnection helps govern the nonlinear saturation of the MRI \cite{sano:2001, fleming:2000}.  Efficient reconnection due to higher resistivity reduces the growth rate of the MRI \cite{fleming:2000}.  The role of reconnection in the nonlinear growth and saturation of the MRI in weakly ionized protostellar disks remains an active area of research.

Observations of protoplanetary disks such as HL Tau often show multiple dark and bright rings in dust continuum emission \cite{alma:2015}, which are often attributed to planet-disk interactions.  Suriano et al.\ present two-dimensional \cite{suriano:2018} and three-dimensional \cite{suriano:2019} simulations to propose reconnection assisted by ambipolar diffusion as the mechanism responsible for the rings and gaps.  In this scenario, the radial current is steepened by ambipolar diffusion into a thin layer near the equatorial plane.  The azimuthal magnetic fields above and below the equatorial plane are in opposite directions.  This configuration provides a magnetic torque that drives fast accretion.  The poloidal field gets dragged in and becomes pinched.  This pinching enables reconnection, which reduces the net poloidal flux.  Plasmas build up in the regions of low poloidal magnetic field strength to form dense rings.  On the other hand, three-dimensional simulations by Hu et al.\ \cite{hu:2019} find that the mass is drained out of the reconnection region in order to form a gap.

\subsection{Dusty astrophysical plasmas\label{dust}}

Models of magnetic reconnection in dusty astrophysical environments such as protoplanetary disks and the neutral phases of the ISM must account for the presence of dust in addition to partial ionization.  
The effects of charged dust particles on reconnection remain largely unknown.  Astrophysical dust particles are not uniform, but rather will have a distribution of sizes \cite{mathis:1977}.  The nature of the size distribution function (including cutoffs) will impact the values of non-ideal coefficients \cite{zhao:2016}.
Reconnection has been proposed as a mechanism for heating chondrules in the proto-solar nebula \cite{levy:1989, eisenhour:1994, lazerson:2010, hubbard:2012}. The effects of dust on reconnection have been investigated in cometary magnetotails \cite{Jovanovic:2005}.

A modification of resistive tearing mode theory for a partially ionized dusty plasma included regimes where dust particle inertia or collisions between neutrals and charged particles were dominant \cite{shukla:2002}.  A subsequent work investigated the stability of thin current sheets in an inhomogeneous dusty plasma \cite{Jovanovic:2002}.  The first three-dimensional simulations of reconnection in a dusty plasma showed dust-neutral relative flow velocities that were about twice as big as the the dust Alfv\'en velocity, as well as heating of dust particles without much heating of the ambient plasma \cite{lazerson:2008}.  A later investigation focused on how a population of dust particles affected the Harris sheet equilibrium, and found an analytical solution for the depleted electron regime \cite{lazerson:2011}.


\section{Experiments on magnetic reconnection in partially ionized plasmas\label{experiments}}

\begin{figure}[ht]
\centering
\includegraphics[width=0.8\textwidth]{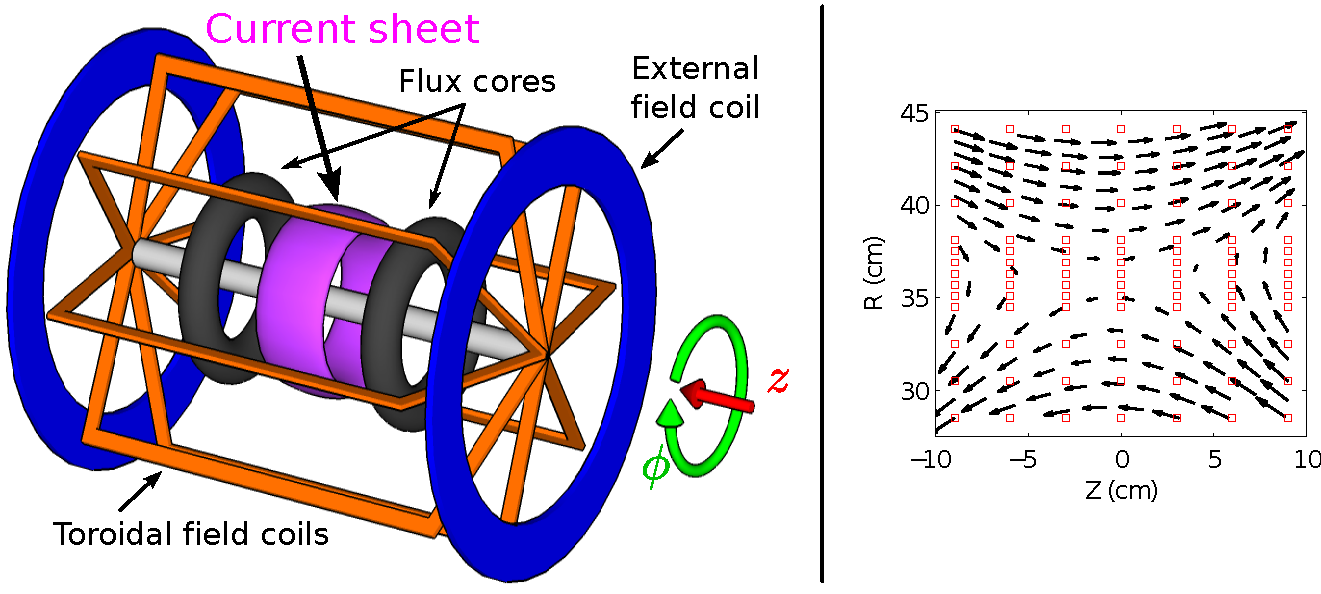}
\caption{\small (left) Schematic drawing of the MRX device.  The gray toroids containt the flux core windings, which produce the potential $X$-point geometry and break down the fill gas.  The blue external field coils modify the initial current sheet position.  The orange toroidal field coils can optionally provide an out-of-plane ``guide field.'' (right) Sample interpolated data from the magnetic vector array. Red squares show actual coil locations.}\label{f:mrxdiagram}
\end{figure}

Laboratory study of magnetic reconnection has become feasible in the last two decades thanks to the significant advances in developing techniques to generate, control and diagnose the process in great detail. However, most of the laboratory studies thus far have focused on magnetic reconnection in fully or mostly fully ionized plasmas where collisionality between electrons and ions has been a primary knob to vary. A summary of these laboratory experiments can be found in \cite{yamada:2010} and will not be repeated here. Instead, a brief summary will be given below only on particular experimental studies on the effects by partial ionization.

\begin{figure}[t]
\centering
\begin{tabular}{ccc}
\hline
\multicolumn{3}{c}{{Ionized Case $\rho_\text{total}/\rho=8$}} \\
\hline
\includegraphics[scale=0.7]{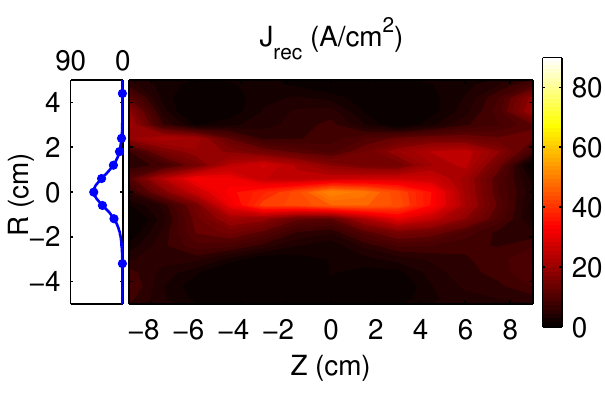} &
\includegraphics[scale=0.7]{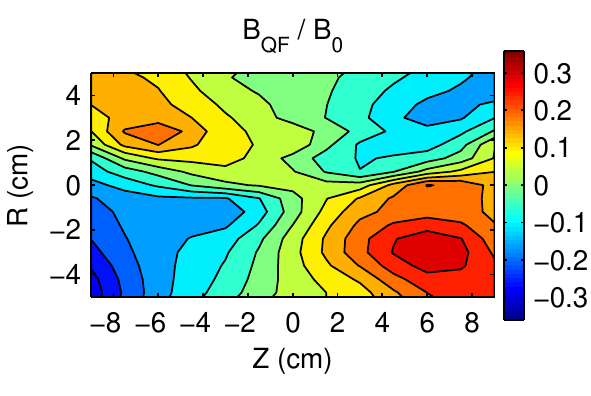} &
\includegraphics[scale=0.7]{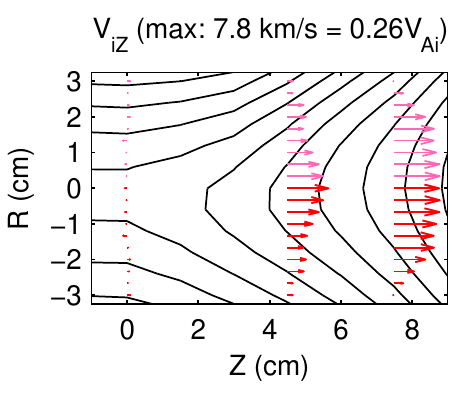} \\
\\
\hline
\multicolumn{3}{c}{{Weakly Ionized Case $\rho_\text{total}/\rho=70$}} \\
\hline
\includegraphics[scale=0.7]{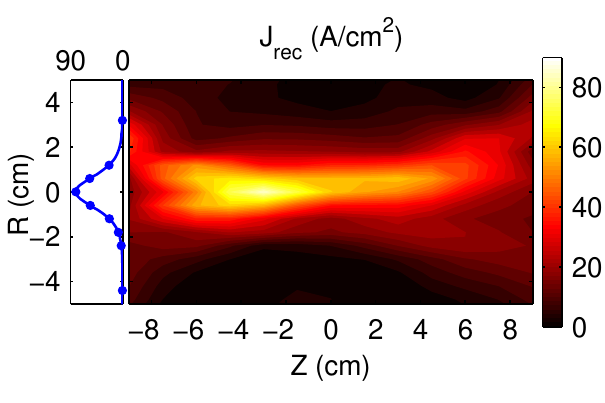} &
\includegraphics[scale=0.7]{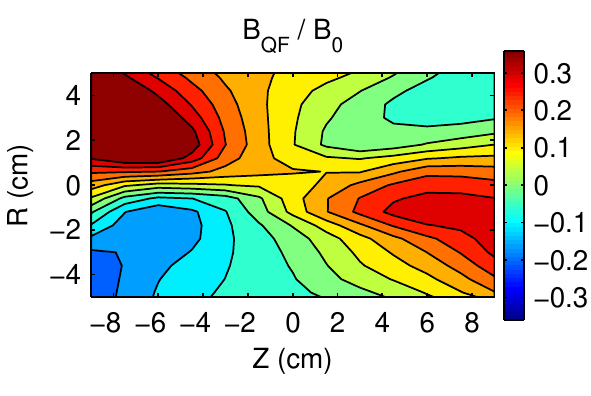} &
\includegraphics[scale=0.7]{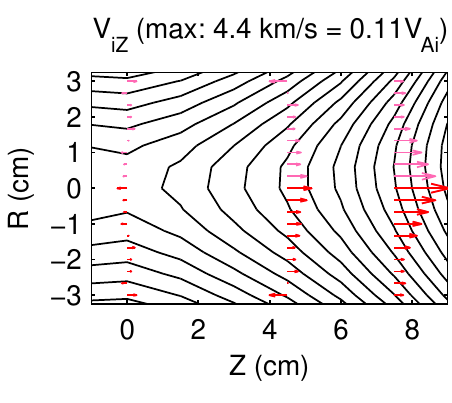} \\
\hline
\end{tabular}
\caption{\small For each ionization case: (left) Reconnection current sheet density. (middle) Out-of-plane magnetic field, scaled to the upstream field. (right) The ion outflow velocity ($Z$ component) with magnetic flux contours. Red arrows are actual measurements, pink arrows are reflections about the current center ($R=0$) to guide the eye.} 
\label{f:jtbt}
\end{figure}

The effects of partial ionization on magnetic reconnection in the Hall regime were systematically studied on MRX~\cite{lawrence:2013} (see Figure~\ref{f:mrxdiagram}) where neutral particles were added to nearly fully ionized plasmas while keeping key plasma parameters, electron density and electron temperature, unchanged. The ionization fraction was lowered to a modest value of $\sim 0.01$. Through this scan, ions in the outflow are decoupled from neutrals at high ionization fractions and become coupled to neutrals at low ionization fractions. However, ions are still largely decoupled from neutrals in the inflow.

\begin{figure}[ht]
\centering
\includegraphics{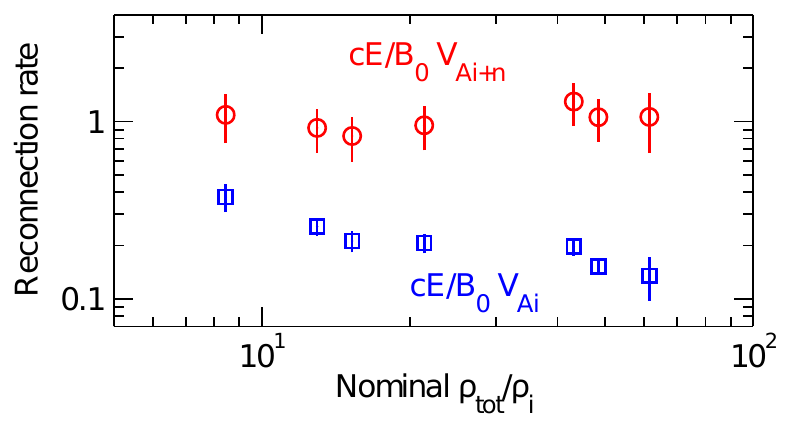}
\caption{\small Reconnection rate normalized to the plasma (blue squares) and bulk (red circles) Alfv\'en speeds, as a function of neutral density ratio.}\label{f:rate}
\end{figure}

Figure~\ref{f:jtbt} shows an example each for ``ionized case'' and for ``weakly ionized case'' for comparisons, for the same electron density of $\sim 2 \times 10^{13} cm^{-3}$ and temperature of $\sim 6$ eV\@. The overall spatial sizes of the current sheet and out-of-plane quadrupole field remain unchanged in contrast to relevant theoretical predictions~\cite{malyshkin:2011}, but consistent with a recent particle simulation~\cite{jara:2019}.
However, the magnitude of the ion outflow was significantly reduced by neutrals in the weakly ionized case without noticeable changes in the outflow channel width. Thus, the ion mass flux and therefore the reconnection rate is reduced when normalized by the ion Alfv\'en speed, see Figure~\ref{f:rate}. We note that if the ion outflow channel width were increased by the effectively increased ion mass $\sqrt{\rho_\text{total}/\rho}$, the reconnection rate then would be independent of the ionization fraction. It is interesting to note that the reconnection rate, when normalized to the bulk Alfv\'en speed (including neutrals), becomes independent of the ionization fraction, although at a much larger value on the order of unity. This was attributed~\cite{lawrence:2013} to the small size of MRX, and therefore it is worth of exploring effects of partial ionization on reconnection at the larger devices such as the Facility for Laboratory Reconnection Experiments (FLARE)~\cite{ji:2011}.

\begin{figure}[ht]
\centering
\includegraphics[width=3in]{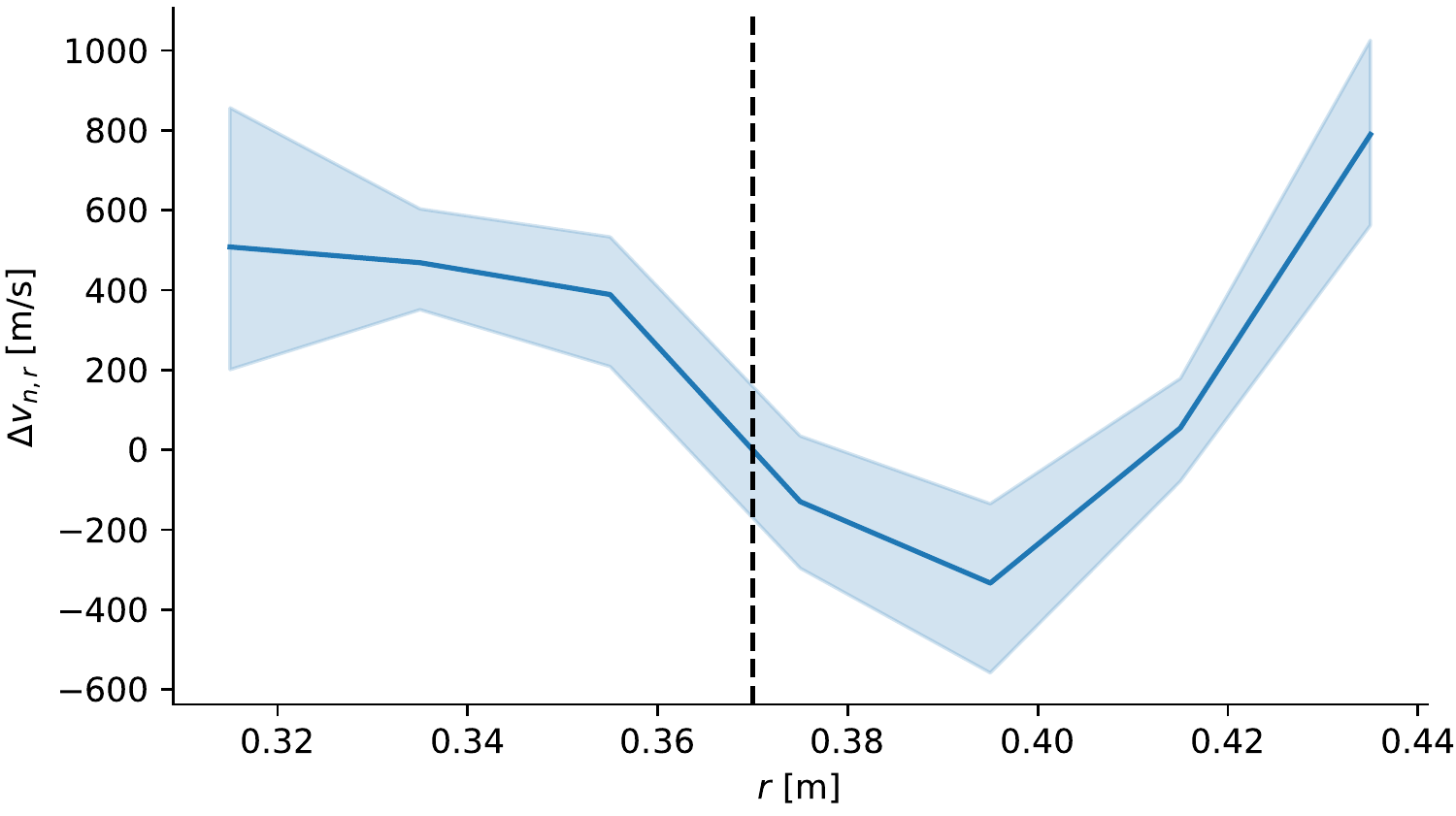}
\caption{\small An example of MRX measurements of the neutral inflow (radial velocity) profile during reconnection in a partially ionized plasma. The errorbars represent the statistical uncertainty as determined by repeated experiments and bootstrapping. The systematic uncertainty is on the order of $\sim 10$ km/s and would lead to an overall shift in the vertical scale. The center of the current sheet (peak of the current density) is shown as a dashed line. Near the center of the current sheet, a small but discernible inflow is measured, while away from the current sheet there is a radial flow shear that is likely due to experimental boundary conditions.}
\label{f:neutral}
\end{figure}

The above experimental research on the effects of partial ionization was indirect without direct measurements of the dynamics of neutral particles. As a first attempt to directly measure neutral particle due to magnetic reconnection in partially ionized plasmas, the Doppler shift of the neutral Helium 587.6 nm complex was measured by an optical probe inserted into MRX. Figure~\ref{f:neutral} shows an example measurement of inflow profile of neutral particles across a reconnecting current sheet. The magnitude of the inflow speed on order of $500$ m/s is about 10\% of local Alfv\'en speed (including neutrals), consistent with the the measured ion flow speed~\cite{lawrence:2013}. The location of the flow reversal is also consistent with the current sheet location, indicating that neutrals are dragged into the reconnecting current sheet through ion-neutral collisions, even as ions are still dynamically decoupled from neutrals in the inflow. We note, however, that this neutral inflow induced by the reconnecting ion inflow can be understood as part of circulation of neutral particles around the reconnecting current sheet~\cite{lawrence:2013}.

A new experiment was recently developed at the University of Tokyo to investigate reconnection in a plasma with an ionization fraction of less than 1\% \cite{takahata:2019, yanai:2018}.  Instead of using inductive torus breakdown, they utilize a rotating magnetic field (RMF) technique to create the plasma configuration.  The time scale of the experiments was limited by available capacitance in the RMF power supply, so the ambipolar diffusion effect was not yet observed.  They anticipate that three-fluid effects will be observed after an upgrade to the power supply.

\section{Summary and future work\label{summary}}

Plasmas are partially ionized in many astrophysical environments. How neutrals affect reconnection is important for understanding the release of magnetic energy in these environments.  Collisions between neutrals and ions alter the resistivity of the plasma, which impacts the dissipation of the magnetic field during the reconnection process.These collisions also contribute to the viscosities and affect momentum and energy transport during reconnection.  Ambipolar diffusion caused by the decoupling between the neutral particles and the ionized components can lead to fast magnetic reconnection when the guide field is zero. Non-equilibrium ionization-recombination effects can also possibly play important roles in the reconnection process.  In the strongly coupled case with weakly ionized plasmas, collisions between neutrals and ions can possibly increase the effective ion inertial length and make the onset of fast collisionless reconnection appear earlier, although this simplified picture has been challenged recently by first-principle PIC simulation~\cite{jara:2019}. The importance of each of these partial ionization effects depends on density, ionization degree, temperature, and magnetic field strength.  Consequently, different effects will be important in different environments.

The temperature in the low solar atmosphere (beneath the upper chromosphere) is usually below $10^4$~K\@. The density stratification from the photosphere to the upper chromosphere changes 6 to 7 orders of magnitude in several thousand km. The plasmas are weakly ionized below the middle chromosphere. The minimum ionization degree is about $10^{-4}$ around the temperature minimum region (about $4200$~K). The hydrogen plasma components are strongly ionized in the upper chromosphere. The topologies of magnetic fields in different reconnection events can be very different, and the strength can vary from dozens of Gauss to several thousands of Gauss or even higher. The magnetic diffusion $\eta$ is mainly contributed by collisions between electrons and ions even in the photosphere. The diffusivities contributed by Hall effect ($\eta_H$) and ambipolar diffusion ($\eta_{AM}$) are much higher than the magnetic diffusion $\eta$ above the middle chromosphere.  High resolution observations at multiple wavelengths from different solar telescopes have recently discovered many small scale reconnection events in the low solar atmosphere, some of which have been studied using MHD simulations.  What we have learned about magnetic reconnection in the partially ionized low solar atmosphere can be summarized as below:

1. Observational results show that many small scale reconnection events frequently happen at different heights in the low solar atmosphere. Emission and absorption at different wavelengths roughly indicate the temperatures and densities in these reconnection events, while blue and red doppler shifts indicate the velocity distributions.  The strength and distributions of magnetic fields in the photosphere can also be roughly measured. However, the magnetic topology, triggering mechanisms, formation and evolution of these reconnection events remain poorly understood because of the limited resolution.

2. The triggering mechanisms and formation of the reconnection events in the low solar atmosphere (e.g., EBs, jets and UV bursts) have been studied by using single-fluid MHD simulations.  Many characteristics (e.g., lifetimes, temperatures, velocity distributions, structures of the images and spectra in different wavelengths) can match well with observations. However, the numerical resistivity or an assumed anomalous resistivity is usually applied to trigger the magnetic reconnection process.  The small scale physics which are very important to understand the reconnection mechanisms and magnetic energy conversion are ignored. These small scale physics can also be crucial to understand some of the observed features which can not be explained by using the present simulation results. For example, some of the synthetic spectra differ greatly from observations. Including more realistic small scale physics in the numerical simulations might drastically change the scenarios and models for explaining the observed reconnection events.  

3. Ambipolar diffusion, which refers to the decoupling of neutrals and charged components, has been studied by using high resolution single-fluid MHD simulations. These simulations show that ambipolar diffusion causes rapid current sheet thinning when a null point topology is present in the lead-up to reconnection.

4. The decoupling of ions and neutrals is more naturally included in a multi-fluid model. The simulations show that the neutral and ion fluids become decoupled upstream from the reconnection site when the reconnection magnetic field is weak and plasma $\beta$ is large. The ions move toward the current sheet, and many neutrals are left behind because of their lower inflow velocity. Recombination and Alfv\'enic outflows quickly remove ions from the reconnection site, leading to a fast reconnection rate independent of Lundquist number before the plasmoid instability appears.

5. When the reconnection magnetic field is strong (about hundreds of Gauss, which is common in solar active regions) and plasma $\beta$ is low, simulations show strong coupling between the ion and neutral flows, and the ionization rate greatly exceeds the recombination rate inside the current sheet. Recombination does not play a significant role in accelerating the reconnection rate, and the plasmoid instability is the main mechanism to lead to fast reconnection in this situation. However, the non-equilibrium ionization-recombination effect is important for studying the temperature variations inside the reconnection region.   

6. The Hall effect is not important below the middle chromosphere. The simulation results based on the reactive multi-fluid plasma-neutral module show that the plasmoid cascading process terminates on scales with length around 100 m and width around 2 m when the reconnection magnetic field around the solar TMR is about $500$~G\@. Since the initially weakly ionized plasmas become strongly ionized inside the current sheet, the ion inertial length becomes much smaller than the width of $2$~m. Therefore, the plasmas are collisional and the magnetic energy is dissipated in the MHD scale in such a reconnection process. However, fast Hall reconnection probably dominates in the upper chromosphere.   

7. Radiative cooling and radiative transfer are very important in the low solar atmosphere, in particular below the middle chromosphere. Numerical results show that radiative cooling contributes to faster current sheet thinning during reconnection with a null point topology. More importantly, radiative cooling modifies the variations and distributions of temperature and ionization degree inside the current sheet.

High resolution multi-fluid simulations based on the radiative and reactive plasma-neutral model with non-equilibrium ionization are the necessary next step in the future modeling of events in the low solar atmosphere such as EBs, UV bursts, and chromospheric jets. Some of the reconnection processes that trigger jets and transition region events happen in the upper chromosphere, so the effects of the Hall term and viscosity on magnetic reconnection there require further numerical study. However, numerically studying these events by including all the small scale physics in large scale simulations from the photosphere to the corona is very challenging. The 4 meter Daniel K.\ Inouye Solar Telescope (DKIST), which recently had its first light, will enable the highest resolution solar observations at many wavelengths when it goes into full service. The proposed 8 meter Chinese Giant Solar Telescope (CGST) is expected to provide very high resolution measurements of the solar vector magnetic field. These high resolution solar telescopes will provide us great opportunities to better understand magnetic reconnection in partially ionized plasmas. The small scale structures (e.g., the plasmoids or flux rope structures) which usually appear in the reconnection region of high resolution numerical simulations can possibly be identified and proved from the future observations. The distributions and variations of temperature, density, velocity and magnetic fields in the reconnection events can possibly be better measured and deduced from the data in multiple wavelengths. These high resolution observations can then be compared with the numerical results or used to restrict the initial and boundary conditions of numerical simulations. The extremely small scale events which have never been observed can possibly be discovered from the future high resolution telescopes. Simultaneous multiwavelength observations of the solar photosphere, chromosphere and corona could help us understand how reconnection events in the low solar atmosphere affect the upper solar atmosphere.  For example, the reconnection process could possibly trigger different kinds of waves that could contribute to the heating of the chromosphere and corona.

Few works have considered how neutrals affect magnetic reconnection in partially ionized astrophysical plasmas. Beside the direct contributions of neutrals on magnetic diffusion and viscosity, the decoupling of ions and neutrals and the recombination of ions have been studied and considered to play important roles to result in current sheet thinning and fast reconnection in the weakly ionized ISM\cite{zweibel:1989, zweibel:2002, zweibel:1997}. Analytical results have shown that the plasmas in molecular clouds and protostellar disks are collisional\cite{ji:2011}, which means that kinetic effects might not be important for fast magnetic reconnection in these environments. Laboratory experiments~\cite{lawrence:2013} and PIC simulations~\cite{jara:2019} show that the Hall reconnection can indeed occur in weakly ionized plasmas, but the reconnection rate will be slowed down when the ionization degree is much smaller than $1\%$. Despite the limitations in achievable parameters, well-controlled and well-diagnosed laboratory experiments in partially ionized plasmas such as those in MRX and the upcoming FLARE or Facility for Laboratory Reconnection Experiments~\cite{ji:2011} should shred new lights into understanding complex dynamics between electromagnetic fields and multiple fluids including neutrals. First-principle simulations using PIC methods can also play a critical role between fluid simulations, laboratory experiments and observations.

A reconnection model in turbulent fluids has been proposed to explain fast reconnection in astrophysical environments \cite{lazarian:1999, lazarian:2014, lazarian:2020}. The authors claim that Hall reconnection is not important in large-scale astrophysical reconnection events and that ambipolar diffusion does not play a significant role for turbulent transport in magnetized fluids with strong turbulence \cite{lazarian:2014}. This turbulent reconnection model and their conclusions need further independent studies to confirm.

Since the differences can be huge on the plasma parameters and strengths of magnetic fields in different astrophysical environments, the main mechanisms which result in fast reconnection rate and plasma heating can be very different. In addition, the plasma parameters and magnetic fields can possibly vary dramatically with time, in which case the dominant fast reconnection mechanism might change with time during magnetic reconnection. Further numerical, experimental and observational studies should focus on the specific time dependent reconnection events.  

\enlargethispage{20pt}


\dataccess{This article has no additional data.}

\aucontribute{L.N.\ wrote most of sections 1--3 and 6.  N.M.\ wrote most of section 4.  H.J.\ and J.J.A.\ wrote most of section 5.  All authors contributed to revisions of the manuscript.}

\competing{We declare we have no competing interests.}

\funding{ This research is supported by NSFC Grants 11973083, 11573064, 11333007, 11303101 and 11403100; the Youth Innovation Promotion Association CAS 2017; the Applied Basic Research of Yunnan Province in China Grant 2018FB009; the Yunnan Ten-Thousand Talents Plan-Young top talents; the key Laboratory of Solar Activity grant KLSA201812; the Strategic Priority Research Program of CAS with grants XDA17040507, QYZDJ-SSWSLH012 and XDA15010900; the project of the Group for Innovation of Yunnan Province grant 2018HC023; the YunnanTen-Thousand Talents Plan-Yunling Scholar Project. H.J.\ and J.J.A.\ acknowledge support by the U.S. Department of Energy Office of Fusion Energy Sciences under Contract No. DE-AC02-09CH11466. N.A.M.\ acknowledges support from the U.S.\ NSF grant 1931388.}

\ack{The authors are grateful to all the referees for their valuable comments and suggestions. L.N.\ would like to thank Dr. Jun Lin for his helpful discussions.}


\vskip2pc

\bibliographystyle{RS}
\bibliography{MR-parti-ionized-RSPA}
\end{document}